\documentclass[fleqn,usenatbib]{mnras}

\usepackage[T1]{fontenc}
\usepackage{ae,aecompl}

\usepackage{graphicx}	
\usepackage{amsmath}	
\usepackage{amssymb}	
\usepackage{blindtext}  
\usepackage{natbib}
\usepackage{rotating,times,graphicx,latexsym}
\usepackage{color}
\usepackage{longtable}
\usepackage{lscape}
\usepackage{lipsum} 
\usepackage{array}
\usepackage{flafter}

\usepackage[pdftex,dvipsnames]{xcolor}  
\usepackage{amsmath}

\newcommand{\hc}{{HOYS}}
\newcommand{\ha}{{$H\alpha$}}
\newcommand{\vc}{{V1598\,Cyg}}

\graphicspath{ {images/} }
\newcommand{\hii}{H{\sc ii}~}
\newcommand{\al}{{$\hat{A}_\lambda$}}
\newcommand{\au}{{$\hat{A}_U$}}
\newcommand{\ab}{{$\hat{A}_B$}}
\newcommand{\av}{{$\hat{A}_V$}}
\newcommand{\ar}{{$\hat{A}_{R_c}$}}
\newcommand{\ai}{{$\hat{A}_{I_c}$}}
\newcommand{\ah}{{$\hat{A}_{H\alpha}$}}

\newcommand{\siu}{{$\sigma_U$}}
\newcommand{\sib}{{$\sigma_B$}}
\newcommand{\siv}{{$\sigma_V$}}
\newcommand{\sir}{{$\sigma_{R_c}$}}
\newcommand{\sii}{{$\sigma_{I_c}$}}
\newcommand{\sih}{{$\sigma_{H\alpha}$}}

\title[\vc]{A survey for variable young stars with small telescopes: III - Warm spots on the active star \vc}

\author[Dirk Froebrich et al.]{Dirk Froebrich$^{1}$\thanks{E-mail: df@star.kent.ac.uk},
Aleks Scholz$^{2}$, Jochen Eisl\"offel$^{3}$, Bringfried Stecklum$^{3}$,
\\
$^{1}$Centre for Astrophysics and Planetary Science, School of Physical Sciences, University of Kent, Canterbury CT2 7NH, UK\\
$^{2}$SUPA, School of Physics and Astronomy, University of St Andrews, North Haugh, St Andrews KY16 9SS, UK\\
$^{3}$Th\"{u}ringer Landessternwarte, Sternwarte 5, 07778 Tautenburg, Germany\\
}

\date{Accepted XXX. Received YYY; in original form ZZZ}

\pubyear{2020}

\begin{document}
\label{firstpage}
\pagerange{\pageref{firstpage}--\pageref{lastpage}}
\maketitle

\begin{abstract}

Magnetic spots on low-mass stars can be traced and characterised using multi-band photometric light curves. Here we analyse an extensive data set for one active star, \vc, a known variable K dwarf which is either pre-main sequence and/or in a close binary system. Our light curve contains 2854 photometric data points, mostly in $V$, $R_c$, $I_c$, but also in $U$, $B$ and \ha, with a total baseline of about 4\,yr, obtained with small telescopes as part of the \hc\ project. We find that \vc\ is a very fast rotator with a period of 0.8246 days and varying amplitudes in all filters, best explained as a signature of strong magnetic activity and spots. We fit the photometric amplitudes in $V$, $R_c$, $I_c$ and use them to estimate spot properties, using a grid-based method that is also propagating uncertainties. We verify the method on a partial data set with high cadence and all five broad-band filters. The method yields spot temperatures and fractional spot coverage with typical uncertainties of 100\,K and 3\,--\,4\,\%, respectively. \vc\ consistently exhibits spots that are a few hundred degrees warmer than the photosphere, most likely indicating that the light curve is dominated by chromospheric plage. The spot activity varies over our observing baseline, with a typical time scale of 0.5\,--\,1\,yr, which we interpret as the typical spot lifetime. Combining our light curve with archival data, we find a six year cycle in the average brightness, that is probably a sign of a magnetic activity cycle. 

\end{abstract}





\section{Introduction}

Magnetic activity is a hallmark of low-mass stars with convective envelopes. The complex of phenomena described as magnetic activity includes transient events like flares and persistent active areas like dark spots in the photosphere or bright plage in the chromosphere. For comprehensive reviews on spot properties in different types of active stars and the methods to study them, see for example \citet{2005LRSP....2....8B} and \citet{2009A&ARv..17..251S}.

One of the traditional tools to study the properties of active areas on (unresolved) stellar surfaces is the analysis of multi-filter light curves. If the rotational modulation of the stellar flux caused by a spot is measured in multiple optical filters, the spot temperature and coverage can be inferred from the wavelength dependence of the amplitude. Moreover, with long-term coverage, the lifetimes of spots, their migration, and the duration of magnetic cycles can be constrained \citep{1995ApJS...97..513H,2008A&A...479..827G}.

In this paper we present a detailed study of an active star observed as part of the \hc\ project \citep{2018MNRAS.478.5091F}. While the primary objective of \hc\ is 'Hunting Outbursting Young Stars' (as evidenced by the acronym), the project delivers long-term, multi-band optical light curves for many types of variable stars. In a previous paper, we used HOYS light curves to study occultations by circumstellar disk material in the young stellar object V\,1490\,Cyg \citep{2020MNRAS.493..184E}. 

Here, we focus on another star in the same region that we found to be consistently variable. \vc\ does not show signs of accretion or circumstellar dust, and the variability is readily explained by magnetic activity. We use the wealth of available \hc\ data to measure spot properties over timescales of years. This includes the development of a robust and automatic process to measure the time evolution of the amplitudes in the various optical broad band filters, including a reliable estimate for the uncertainties. Furthermore, we develop a grid-based fitting procedure for spot properties and their uncertainties based on these amplitudes. These computational routines are tested and evaluated for \vc, but can readily be applied to the vast number of periodic variables (esp. young stars) whose multi-filter, high cadence, long term light curves are available from the \hc\ project.

Our paper is organised as follows. We review the literature for \vc\ in Sect.\,\ref{target}. In Sect.\,\ref{data} we describe the \hc\ data obtained for the star. We detail the general light curve properties in Sect.\,\ref{variability} and describe our modelling of spot properties in Sect.\,\ref{spotmodelling}. In Sect.\,\ref{discussion} we discuss our findings.

\section{\vc\ in the literature}
\label{target}

\vc\ is known as a variable star in the literature (named 'B14' prior to being included in the list of variable stars, \citet{1978IBVS.1414....1K}), but the nature of the star and its variability has not been investigated yet in detail. Flares have been reported by \citet{1974IBVS..938....1T} and \citet{1987IBVS.2981....1R}, with U-band amplitudes of 1.7 and 2.6\,mag. A recent long-term study by  \citet{2018RAA....18..137I} found 'no significant photometric variability', although inspection of their light curve does indicate weak long-term trends. Their sparse sampling does not allow the detection of flares or other short-term variations on timescales of hours or days. As shown below, we find a period shorter than one day, which requires high cadence observations to be distinguished from white noise.

On the sky, \vc\ is located in the IC\,5070 \hii region, which is associated with the Pelican Nebula, a well known star forming region close to the Galactic plane. The star is found near (slightly south) of the sub-cluster 5 identified in the Spitzer/IRAC survey by \citet{2009ApJ...697..787G}, but is not listed in their catalogue of YSOs (presumably due to lack of infrared excess, see below). The Gaia DR2 parallax of $2.44 \pm 0.02$\,mas \citep{2018yCat.1345....0G}, places the star at a distance of $405\pm 3$\,pc \citep{2018AJ....156...58B}, whereas IC\,5070 has a significantly larger distance, recently estimated as 870\,$^{+70}_{-55}$\,pc \citep{2020MNRAS.493..184E}. Thus, \vc\ is a foreground star and does not belong to the star forming region.

Photometry for this star is available in many catalogues. It is listed at $V=14.1$\,mag \citep{2016yCat.2336....0H}, $G=13.8$\,mag \citep{2018yCat.1345....0G}, $J=12.2$\,mag, $H=11.8$\,mag, and $K=11.5$\,mag \citep{2013ASSP...37..279L}, with 0.1\,--\,0.2\,mag deviations between catalogues and epochs, presumably due to long-term variability. For example, the 2MASS $H$ band magnitude is $11.6$\,mag \citep{2006AJ....131.1163S}. The combination of $V-K$ and $J-K$ colours is consistent with a mid-K dwarf star without significant reddening \citep{1988PASP..100.1134B}. Based on the $V-K$, $J-H$ and $H-K$ colours, it is a K3\,--\,K4 star with effective temperature of 4700\,K and an approximate mass of 0.7\,M$_{\odot}$ \citep{2013ApJS..208....9P}\footnote{\tt https://www.pas.rochester.edu/$\sim$emamajek/EEM$\_$dwarf$\_$UBVIJHK$\_$colors$\_$Teff.txt}. In the TESS input catalogue, the object is listed as a 0.8\,M$_{\odot}$ dwarf star \citep{2019AJ....158..138S}.

In the mid-infrared, the star is listed at $W1=11.35$\,mag, $W2=11.36$\,mag and $W3=11.42$\,mag in ALLWISE \citep{2013yCat.2328....0C}. The negligible colours in these bands at 3\,--\,12$\,\mu$m preclude the existence of warm dust in a circumstellar disk. In IPHAS, \vc\ is catalogued with $r=13.82$\,mag, $i=13.18$\,mag and \ha\,$=13.44$\,mag \citep{2014MNRAS.444.3230B}. Comparing with the simulated colours for emission line stars tabled in \citet{2011MNRAS.415..103B}, this is consistent with an inactive or weakly active K-star without significant reddening. 

\begin{figure}
\centering
\includegraphics[width=\columnwidth]{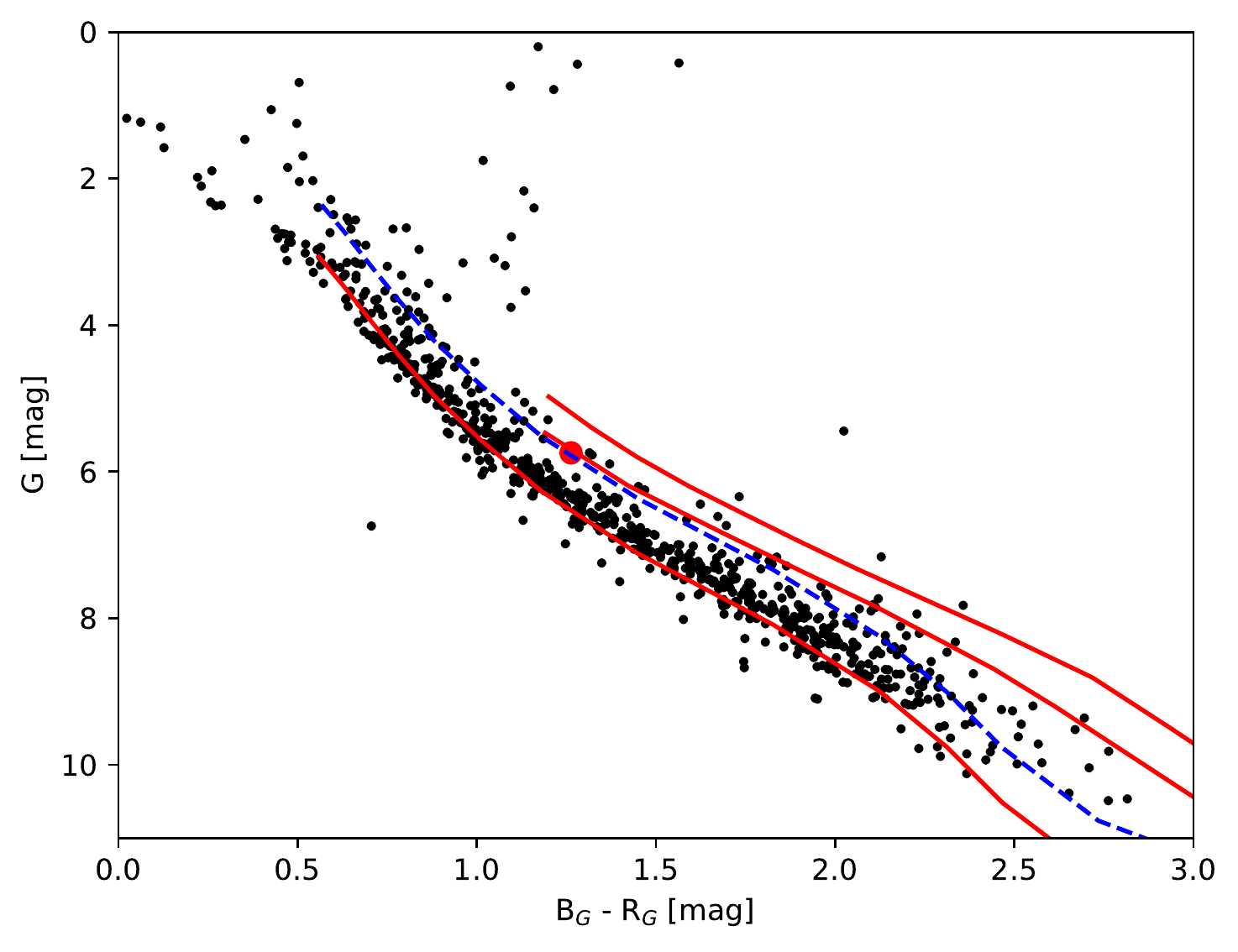}
\caption{\label{region_hrd} Gaia HRD showing the absolute Gmag vs. the B$_G$-R$_G$ colour. Only stars within one degree from \vc, with a parallax SNR better than 10, a parallax larger than 2\,mas and photometric uncertainties smaller than 0.01\,mag are shown. \vc\ is indicated by the larger, red circle. The equal-mass binary sequence is shown as dashed blue line. The 10, 20\,Myr and 1\,Gyr, solar metallicity isochrones by \citet{2010A&A...513A..19F} are over plotted as red solid lines.}
\end{figure}

In Fig.\,\ref{region_hrd} we show a Hertzsprung-Russel Diagram (HRD) of sources within 1\,deg from \vc, generated using Gaia DR2. Only sources with a parallax of more than 2\,mas (to exclude young stars in IC\,5070), a parallax signal to noise ratio (SNR) above 10, and photometric uncertainties in Gaia G/B/R below 0.01\,mag are included. The plot also contains 10, 20\,Myr and 1\,Gyr solar metallicity isochrones from \citet{2010A&A...513A..19F}\footnote{\tt https://phoenix.ens-lyon.fr/Grids/BT-Settl/CIFIST2011bc/ISOCHRONES/} as solid red lines as well as the sequence for equal mass binaries as dashed blue line. From this diagram it is apparent that \vc\ (marked as the larger red circle) is located about 0.7\,mag above the Main Sequence. Displacement above the main sequence like this can either be a sign of youth, binarity, or super-solar metallicity. It seems unlikely that the metallicity alone can explain the entire 0.7\,mag off-set as very high values would be needed (e.g. \citet{2002MNRAS.335.1147K}). But youth (age of 20\,Myr) or binarity (mass ratio of one) on their own can create such a shift. \vc\ can hence be an object at those extremes, or a source where a combination of these factors contributes to the off-set from the main sequence. It hence has a minimum age of 20\,Myr, a possible companion and potential super-solar metallicity. 

There is insufficient observational data available at the moment to distinguish between these possibilities. In Gaia DR2, the source has approximately the typical proper motion of the field population and there is no indication of any clustering in proper motion space indicating a group of co-moving (young) objects associated with \vc. This remains true when selecting only stars with a parallax similar to \vc. Thus, there is no kinematic evidence that \vc\ is a pre-main sequence star. On the other hand, the star is astrometrically 'well behaved' in Gaia DR2, without astrometric excess noise and with a re-normalised unit weight error (RUWE\footnote{\tt http://www.rssd.esa.int/doc$\_$fetch.php?id=3757412}) of 1.21, i.e. there is no sign for a companion in the astrometry. If it is a binary, it is either a very close ($<0.1$\,AU) or a wide ($>10$\,AU) system \citep{2020arXiv200305467B}. Figure\,\ref{region_hrd} also indicates that there is a number of potential pre-main sequence stars in the field.

\section{Observational Data}\label{data}

\subsection{\hc\ Data}

The \vc\ data analysed in this paper has been obtained as part of the \hc\ project \citep{2018MNRAS.478.5091F}. \hc\ is a citizen science project;  professional and amateur observers obtain optical ($U$, $B$, $V$, $R_c$, \ha, $I_c$) images of 25 nearby young clusters and star forming regions with the purpose of constructing long-term multi-filter light curves of young stars. At the time of writing, the project has 67 active participants and has accumulated 29000 images with about 133 million usable brightness measurements of stars. 

For the purpose of this paper we have extracted 2854 brightness measurements of \vc\ from the \hc\ database for analysis. They cover a total of 1545\,d (just over 4\,yr) from JD\,=\,2457463 to 2459007. There are 84 measurements in $U$, 390 in $B$, 889 in $V$, 848 in $R_c$, 74 in \ha, and 569 in $I_c$. A large fraction of the data is concentrated in a high cadence campaign that was conducted between  MJD\,=\,58330 and 58400. This campaign monitored the source V\,1490\,Cyg in the same field \citep{2020MNRAS.493..184E} but also generated valuable data for \vc. During this high cadence period there are 77 measurements in $U$, 269 in $B$, 408 in $V$, 402 in $R_c$, 66 in \ha, and 210 in $I_c$.

\begin{figure}
\centering
\includegraphics[width=\columnwidth]{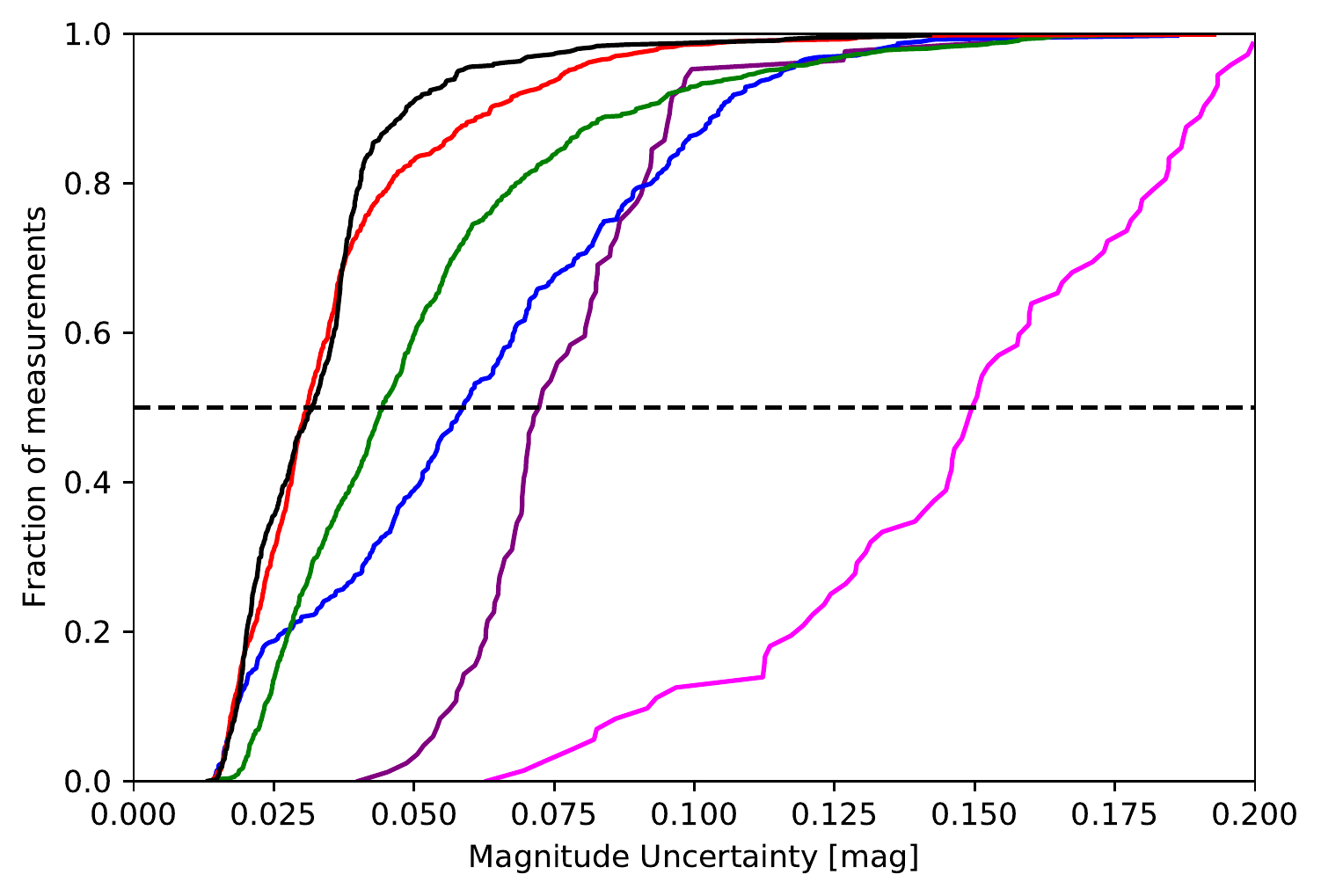}
\caption{\label{uncert} Cumulative distributions of the photometric uncertainties of the \hc\ data of \vc\ after the colour calibration procedure for all filters, colour coded as: $U$ purple, $B$ blue, $V$ green, $R_c$ red, \ha\ magenta, $I_c$ black. The dashed horizontal line allows to estimate the median uncertainty.}
\end{figure}

The \hc\ data is calibrated in two separate steps. At first, the magnitudes are calibrated into a reference system for each of the filters \citep{2018MNRAS.478.5091F}. As many of the \hc\ images are taken through slightly different filters, we have developed an additional colour correction of the photometry \citep{2020MNRAS.493..184E}. This procedure identifies non-variable stars in each \hc\ target field and uses their colours and magnitudes to determine colour terms for each individual image. Afterwards the colour correction is applied to all stars. In \citet{2020MNRAS.493..184E} we have shown that this results in typical photometric uncertainties of the broad band photometry of a few percent. In Fig.\,\ref{uncert} we show the normalised cumulative distribution functions for the photometry of \vc. The median uncertainties in the magnitude measurements are 0.072\,mag in $U$, 0.059\,mag in $B$, 0.044\,mag in $V$, 0.031\,mag in $R_c$, 0.150\,mag in \ha, and 0.032\,mag in $I_c$. 

\subsection{Supplementary long term photometric data}\label{aux_data}

\begin{figure*}
\centering
\includegraphics[width=2.0\columnwidth]{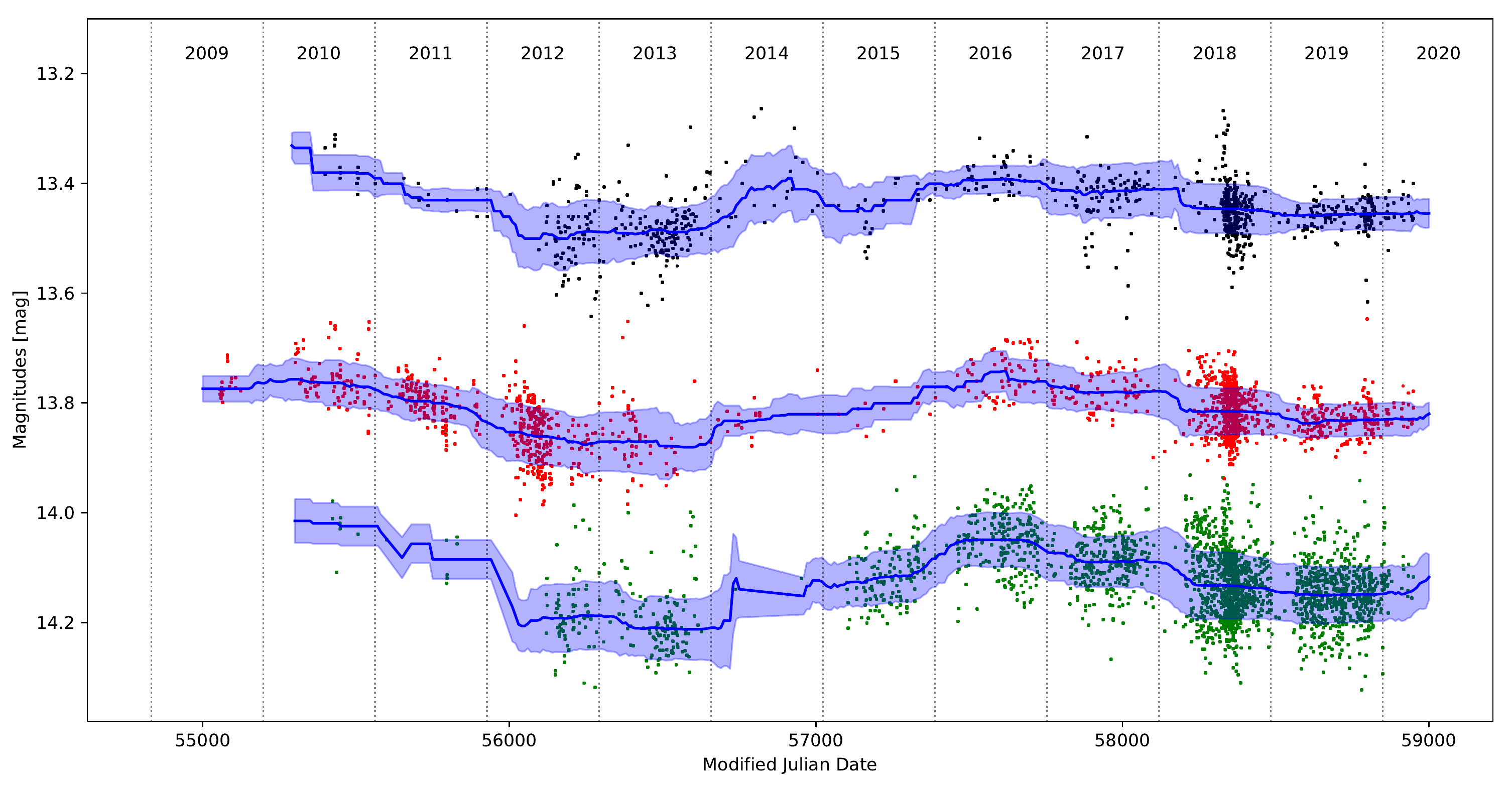}
\caption{\label{longtermlc} Long term light curve of \vc\ for the $V$ (bottom, green), $R_c$ (middle, red) and $I_c$ (top, black) filters. We have included all data from the \hc\ project, as well as all data from ASAS-SN, PANSTARRS, PTF, ZTF, AAVSO and \citet{2018RAA....18..137I}. All the non-\hc\ data have been off-set into the \hc\ photometric system using the median magnitudes in overlapping years. The solid lines represent the running median magnitude determined for a $\pm$\,150\,d window and the shaded areas are the one sigma scatter from the running median.}
\end{figure*}

In order to investigate the long term photometric behaviour of \vc\ beyond the current data available from \hc, we have collected auxiliary data from a number of resources. These include the following: i) Data from the All-Sky Automated Survey for Supernovae (ASAS-SN, \citet{2014ApJ...788...48S}, \citet{2017PASP..129j4502K}) taken in the $V$ and $g$ filters. ii) Data from the Panoramic Survey Telescope and Rapid Response System (Pan-STARRS, \citet{2016arXiv161205560C}, \citet{2016arXiv161205243F}) taken in the $g$, $r$ and $i$ filters. iii) Data from the Intermediate Palomar Transient Factory (iPTF, \citet{2009PASP..121.1334R}, \citet{2009PASP..121.1395L}) taken in the $g$ and $r'$ filters. iv) Data from the Zwicky Transient Facility (ZTF, \citet{2019PASP..131a8002B}, \citet{2019PASP..131a8003M}) taken in the $g'$ and $R$ filters. v) All photometric data of the source from \citet{2018RAA....18..137I} taken in the $V$, $R_c$ and $I_c$ filters. vi) All photometric data available in the American Association of Variable Star Observers (AAVSO) database\footnote{\tt https://www.aavso.org} in the $V$ and $I$ filters - there are only five $R$ measurements. 

We have also checked the data from the Transiting Exoplanet Survey Satellite (TESS). The star is at the faint limit of TESS and the light is influenced by at least one, nearby star of similar brightness, so that these data are not useful for our purposes. Furthermore, the Kepler space telescope has not observed this field. In order to compare the available supplementary data which has been taken in part in different filters, we have used the median magnitudes in overlapping years to shift the magnitudes into the photometric system used in \hc. Note that this supplementary data is only used in the analysis of the long term light curve properties (Sect. \ref{general}). All other analysis is solely based on the \hc\ data set.

In the following we provide a general discussion of the light curve for our target, prior to modelling. The full long term light curves in $V$, $R_c$, $I_c$ filters with all \hc\ and auxiliary data (see Sect.\,\ref{aux_data}) are displayed in Fig.\,\ref{longtermlc}. For the purpose of better visibility of long term trends, we overlaid a running median determined for a $\pm$\,150\,d window and the one sigma scatter from the median. The available data now spans about 10\,yr of more or less continuous coverage in those three filters. In the other filters ($U$, $B$, \ha) the data is mostly concentrated in 2017\,--\,19, and hence not shown in Fig.\,\ref{longtermlc}.

\section{The variability of \vc}\label{variability}

\subsection{General phenomenology and long term trends}
\label{general}

As can be seen in Fig.\,\ref{longtermlc}, the long term median brightness of \vc\ varies at the 0.1\,mag level in $I_c$ and $R_c$ and about 0.15\,mag in $V$. When high-cadence monitoring is available, our light curve displays an obvious periodicity on timescales of about a day, but the amplitude of that period changes over time (see below). On longer timescales of years, there are gradual increases and decreases with apparent maxima occurring in 2010 and 2016, and minima in 2013 and 2019. This suggests a long-term period of about 6.0\,--\,6.5 years. The general trends in the light curve are mirrored in all filters, when simultaneous coverage is available. The extent of the variability changes slightly with filter, with increasing amplitude for shorter wavelengths. 

Our light curve does not show clear evidence for flaring, although some positive outlier data points could in principle be caused by a flare event.  We note that in most nights our sampling would not be sufficient to resolve magnetic flares that typically last only a few hours. 

\subsection{Period of \vc}\label{period_star}

\begin{figure}
\centering
\includegraphics[width=\columnwidth]{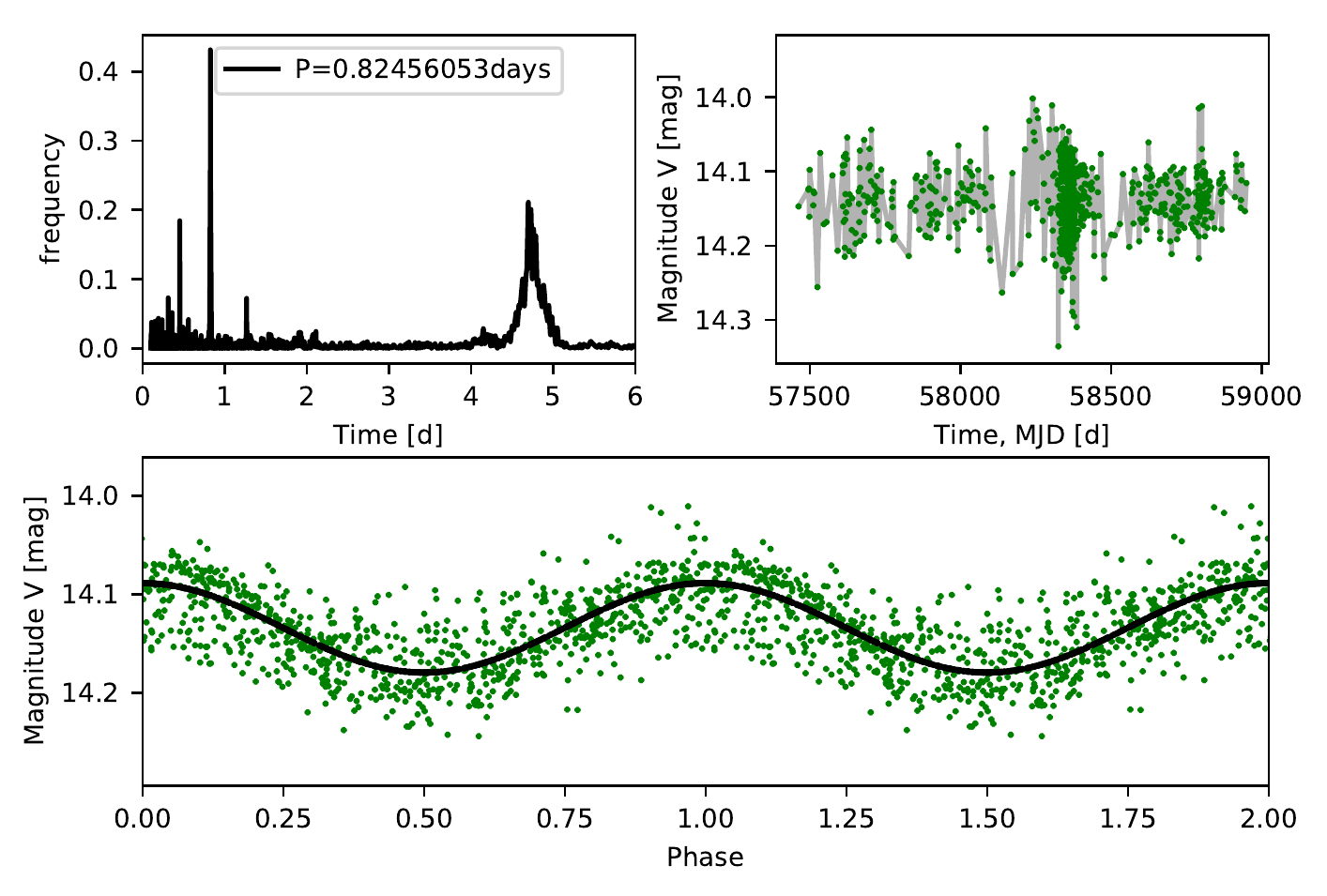}
\caption{\label{lombscarcle_V} Lomb-Scargle periodogram (top left), light curve (top right) and phased light curve (bottom) in V of \vc\ from the \hc\ data.}
\end{figure}

For the analysis of the variability we need to determine the period and amplitude of the variations for the light curve in a particular filter, either for the entire light curve or for parts of it. In a first step we need to remove any long term trends, which are evident in Fig.\,\ref{longtermlc}, by subtracting the running median from the light curve. This running median needs to be determined over a time window much larger than the period, smaller than the long term photometric trends, and ensure there are sufficient data points to determine the median reliably at each point in time. We hence chose a $\pm$\,50\,d window for the running median to be subtracted from the \hc\ light curve of \vc. The median magnitude in each filter over the entire light curve was then added again to ensure any plots show the correct median brightness. 

To verify the presence of a period, we computed  Lomb-Scargle periodograms \citep{1982ApJ...263..835S} over the entire \hc\ light curve for each filter, using the {\tt astropy.stats.LombScargle} implementation. The periodogram shows three clear peaks in all filters, including $U$ and \ha\ (see Fig.\,\ref{lombscarcle_V} for the $V$ band - all other periodograms are shown in Appendix\,\ref{lombscarcle_all}). The strongest peak has a period of 0.82\,d and is very narrow. The two remaining peaks are easily explained as aliases caused by the strong clumping of the data points in intervals of about one day (see \citet{2018ApJS..236...16V} for a discussion). One occurs at 0.45\,d, corresponding to a frequency $f_1 = f_{\mathrm{true}} + 1$ and one occurs at $f_2 =f_{\mathrm{true}} - 1$. The latter peak at a period of approximately 4.7\,d corresponds to the time it takes to sample the same part of the 0.82\,d period at our typical one day cadence.

In the following, we only consider the main period of 0.82\,d. When measured in all the individual filters over the entire \hc\ data set, the median period is $0.8246 \pm 0.0004$\,d, i.e. 19\,h 47\,m 25\,s $\pm$ 35\,s. The fact that this period is consistently found in the entire data set and in all bands suggests that it is the rotation period of our target. If \vc\ is in fact a binary star in tidally locked rotation (see below), the rotation period also constitutes the orbital period. The folded light curve in the bottom panel of Fig.\,\ref{lombscarcle_V} indicates that the general shape of the variability can be described by a sine function (over plotted in the figure). The specific shape of the light curve modulated by spots depends on a number of properties such as the inclination of the stellar rotation axis, the spot latitude, size and shape, as well as limb darkening on the star (see for example \citet{2017A&A...599A...1S}). There is some significant scatter of the data away from the fitted sine function visible in Fig.\,\ref{lombscarcle_V}. This is in part caused by photometric outliers, but can mostly be attributed to a variability in the amplitude over time (see below). 

\subsection{Amplitudes of the \vc\ variability}\label{ampfitting}

Given the shape of the folded light curve in Fig.\,\ref{lombscarcle_V} we fit a sine function to the light curve to determine the amplitude. The sine function has the form:

\begin{equation}\label{sinfunc}
m^\lambda(t_\lambda) = m_0^\lambda + \hat{A}_\lambda \sin\left(\theta_\lambda + 2 \pi \frac{t_\lambda}{P_\lambda}\right).
\end{equation}

Here $P_\lambda$ is the period, $\theta_\lambda$ the phase offset, \al\ the amplitude, $t_\lambda$ the time of the observation, and $m^\lambda(t_\lambda)$ the magnitude as a function of time. Note that the $m^\lambda(t_\lambda)$ values are already corrected for long term trends using a running median. The value of $m_0^\lambda$ corresponds to the median magnitude over the part of the fitted light curve. All the parameters depend on the wavelength of the observation (i.e.  $\lambda$ represents the filter used). 

The fitting is carried out using a least square optimisation procedure. This requires an initial guess for all parameters. For the period we use the median period over all filters estimated from the periodograms (0.8246\,d). For the amplitude we use $2 \sqrt{2}$ times the $rms$ of the magnitudes, which is the theoretical value for the amplitude of a homogeneously sampled sine shaped light curve. For $m_0^\lambda$ we use the median magnitude of the star. There are ways to estimate an initial value for $\theta_\lambda$, but they fail for our under sampled data, caused by the very short period and typical cadence. Hence, we choose a random start value between 0 and $2 \pi$. This works well, but fails to generate a correct fit in about one percent of the cases. Thus, we repeat the fitting procedure ten times with homogeneously distributed start values for $\theta_\lambda$ between 0 and $2 \pi$. The parameters resulting in the fit with the lowest $rms$ are then chosen as the best outcome. During each fit, photometric data points that are more than 3$\sigma$ away from the best fit are iteratively removed from the input data.

We aim to also derive a realistic estimate of the uncertainties of all parameters, in particular for the amplitudes, from this fitting procedure. To achieve that, we repeat the above described fit 30 times. Each time, the individual photometry values in the light curve are randomly varied according to their photometric uncertainties estimated by the colour calibration (see Sect.\,\ref{data}). In other words we add a random value to the magnitude that is drawn from a Gaussian distribution with a mean of zero and a sigma equal to the photometric uncertainty of the data point. We then adopt the mean of the parameters from the 30 fits as the final result, and the $rms$ of the values as their one sigma uncertainty.

We have performed several tests to check the validity of the method. The \vc\ data was used to mimic the same cadence and photometric uncertainty distribution, but the magnitudes were altered to a sine function with known parameters according to Eq.\,\ref{sinfunc}. We then check if the sine parameters can be recovered. The very short period of 0.8246\,d means that our data is typically under sampled. This means that generally the fit is not able to reliably recover periods and phase off-sets if the initial guess for the period is not correct. However, at the same time, all amplitude values are recovered extremely reliably, down to values as low as 0.02\,mag. The typical uncertainties of all the recovered amplitudes are of the order of 0.01\,mag throughout the entire data set. This also holds if the real period is different from the initial guess in the fit by several times the period uncertainty of 0.0004\,d. Thus, our data and fitting procedure allow us to determine reliably and accurately the amplitudes of the variability in \vc. However, due to the short period and our sampling we are unable to identify potential period changes and phase shifts, which would hint to differential rotation and/or spot movement. The latter might be possible to obtain for longer period variables in the \hc\ sample.

\begin{table*}
\caption{\label{tab_amplitudes} Amplitudes and uncertainties in magnitudes measured for \vc\ in all filters. Listed are the average values across the entire light curve (1st row) and the high cadence part of the data (2nd row). All values are in magnitudes.}
\centering
\begin{tabular}{l|cc|cc|cc|cc|cc|cc}
Duration & \au\ & \siu\ & \ab\ & \sib\ & \av\ & \siv\ & \ar\ & \sir\ & \ah\ & \sih\ & \ai\ & \sii\ \smallskip \\   \hline
Entire light curve & 0.088 & 0.0098 & 0.068 & 0.0044 & 0.044 & 0.0019 & 0.035 & 0.0015 & 0.084 & 0.0200 & 0.016 & 0.0021 \\
High cadence data & 0.089 & 0.0116 & 0.076 & 0.0043 & 0.061 & 0.0023 & 0.050 & 0.0023 & 0.084 & 0.0220 & 0.032 & 0.0027 \\
\end{tabular}
\end{table*}

To obtain an overall picture of the variability in \vc\ we have used our procedure to measure the amplitudes in all filters averaged over the full data set, as well as just for the high cadence part. The measured amplitudes and uncertainties are summarised in Table\,\ref{tab_amplitudes}. It is evident that the amplitudes increase with decreasing wavelength, a signature expected for spots in general. The elevated amplitude in \ha\ compared to $R$ strongly suggests that the variability is associated with magnetic activity. The long-term period of $\sim$6\,yr may correspond to a magnetic activity cycle. Amplitudes in $U$ and \ha\ are almost identical for the entire light curve and the high cadence data since almost all the photometry in those filters has been taken in the latter part of the data.

Typical rotation periods for early K main-sequence stars are between 10 and 30\,d, depending on age; typical amplitudes are in the mmag regime \citep{2014ApJS..211...24M}. Thus, compared to main-sequence siblings, \vc\ is a very fast rotator with strong magnetic activity. Fast rotating low-mass stars are either not on the main sequence yet and still in the process of spin-up \citep{2015A&A...577A..98G}, or they are close binaries \citep{2019ApJ...871..174S} with strong tidal interaction \citep{2019ApJ...881...88F}. 

\section{Spot Modelling}\label{spotmodelling}

In this section we describe how we inferred spot temperature $T_S$ and filling factor ($f$, percentage of surface coverage on one hemisphere) from the photometric amplitudes measured from the light curves. The underlying assumption is that the variability is caused primarily by one large spot or spot group, which can be either warmer or cooler than the stellar photosphere. The light curve amplitude \al\ is then calculated from the flux ratio of the un-spotted ($F^\lambda_0$) and spotted surface ($F^\lambda_S$) as:

\begin{equation}\label{eq_amp}
    \hat{A}_\lambda = -1.25 \times \log\left( \frac{F^\lambda_0}{F^\lambda_S}\right) = -1.25 \times \log\left( \frac{F^\lambda_0}{F^\lambda_0(1-f) + F^\lambda(T_S)f}\right)
\end{equation}

For a hot spot, the light curve maximum would correspond to the hot spot being on the hemisphere pointing towards the observer; for a cool spot it is the minimum. As we do not know a priori whether the spots are hot or cold, we use only the absolute values of \al\ when comparing the model amplitudes with the measured values. The factor of 1.25 (instead of 2.5) in Eq.\,\ref{eq_amp} is to allow direct comparison with the sine function amplitudes obtained from the light curve data in Eq.\,\ref{sinfunc}.

To estimate the fluxes $F^\lambda$ we use the PHOENIX model stellar spectra by \citet{2013A&A...553A...6H} provided by the {\tt astropy.pysynphot} module. We fix $\log{g}=4$ and $[M/H]=0.0$.  These are realistic values for our star, and small changes in those parameters lead to only negligible changes in the estimated spot properties. The model spectra are convolved with the $U$, $B$, $V$, $R_c$ and $I_c$ filter curves from {\tt speclite.filters} to determine the magnitudes. 

To fit the amplitudes estimated from the light curves for a subset (or all) of the filters, there are in principle two ways: i) use an optimisation, for example with a least-square method; ii) use a grid of models. We utilise both methods for different parts of our analysis.

\subsection{Estimating the stellar surface temperature}\label{findtstar}

As discussed in Sect.\,\ref{target}, the broad band optical and near infrared colours suggest an early K-type star, either on the pre-main sequence and/or in a close binary system. This implies that the effective surface temperature of the object should be of the order of 4700\,K. There are some uncertainties associated with this value, in particular, it is assuming a single main-sequence star and does not take into account variability. Thus, we aim to determine the stellar temperature from our data set, using the best fit to the observed amplitudes. For this exercise, we use the part of the light curve that includes measurements in all five broad band filters during the high cadence part of the data ( between MJD\,=\,58330 to 58400).

\begin{figure}
\centering
\includegraphics[width=\columnwidth]{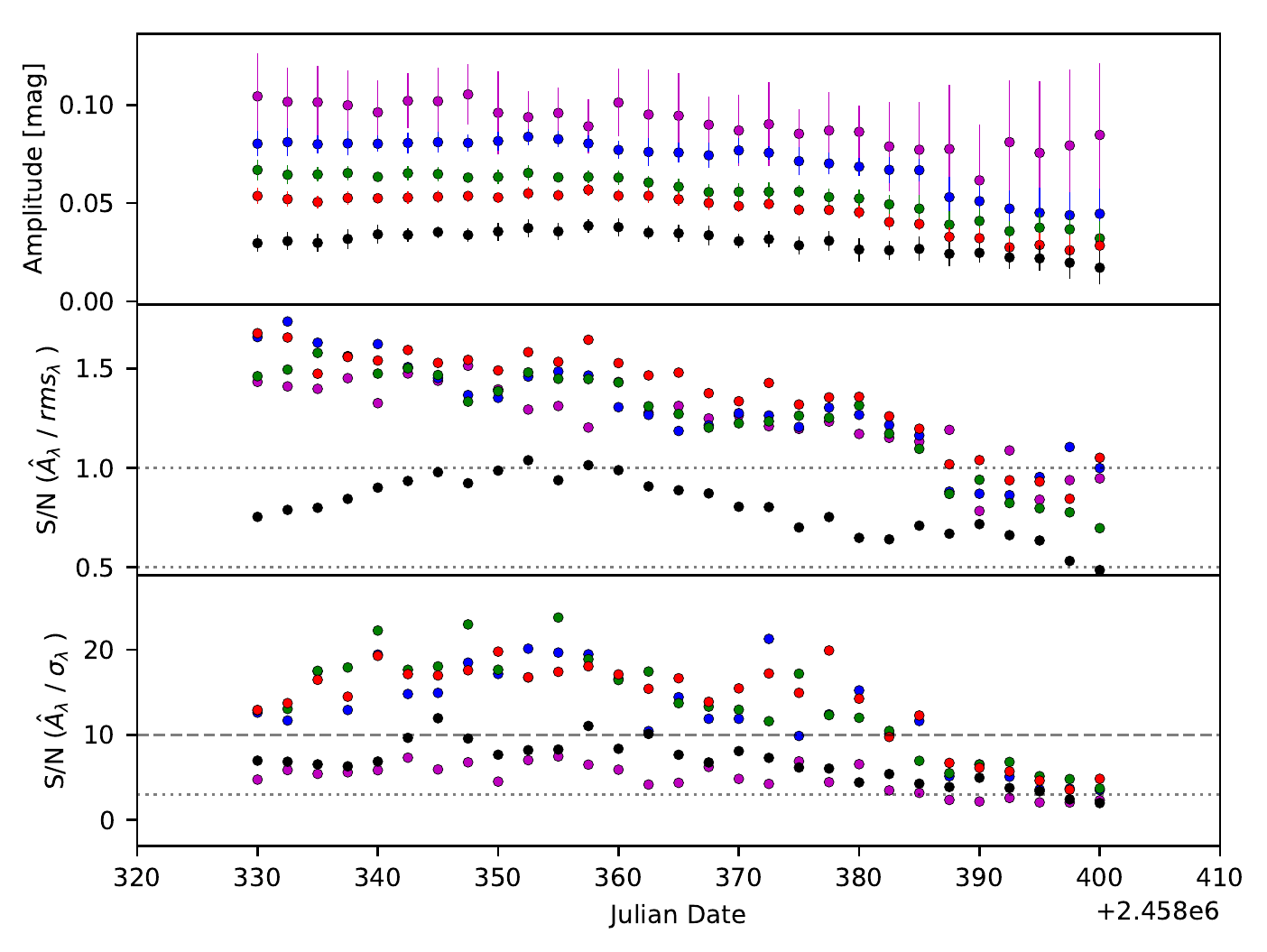} \\
\includegraphics[width=\columnwidth]{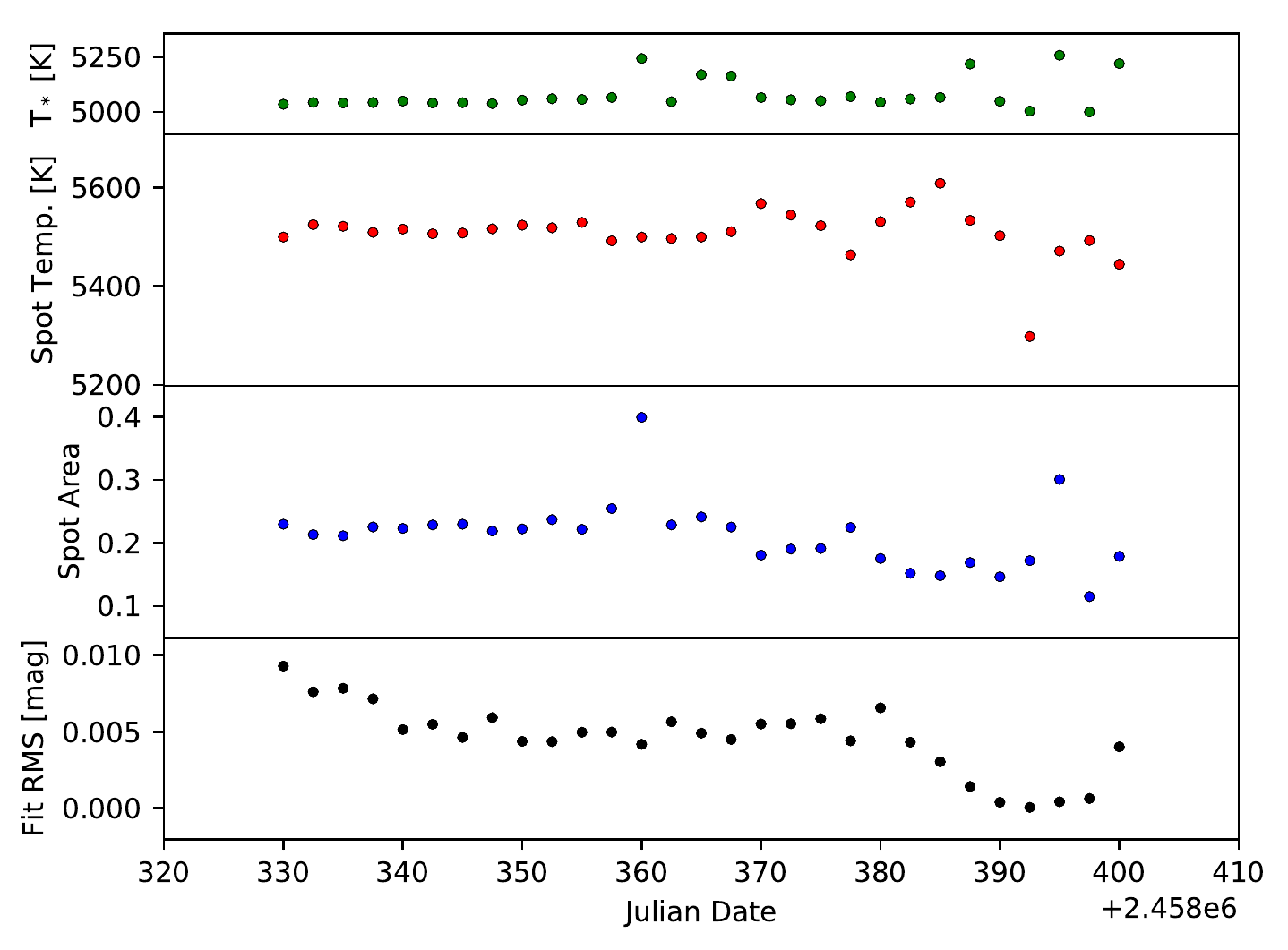}
\caption{\label{fig_test5} {\bf Top:} Determined properties of the \vc\ light curve during the high cadence data in all broad band filters. At each epoch, we include a $\pm$\,15\,d window of data. The first panel shows the light curve amplitudes and their uncertainties. The second panel shows the ratio of the amplitudes and the $rms$ of the photometry after subtraction of the sine fit. In the third panel we show the ratio of amplitudes and their determined uncertainties. {\bf Bottom:} Determined spot properties using the least square optimisation. The first panel shows the star temperature, the second the spot temperature, the third the spot coverage and the forth the $rms$ of the fitted amplitudes.}
\end{figure}

The much higher density of data during this phase, allows us to determine the amplitudes in all five broad band filters in a much smaller time window, i.e. higher temporal resolution. We have chosen to use all photometry within a $\pm$15\,d window at 29 epochs, spaced every 2.5\,d. Subsequently we determine the best-fitting surface parameters using the least square optimisation method. In Fig.\,\ref{fig_test5} we show the determined broad band amplitudes and the determined spot and star properties (stellar temperature, spot temperature and coverage). With the exceptions of a few epochs, the best fit parameter values are consistent. In particular, a stellar temperature of approximately 5050\,K leads to the best fit in most cases. We are hence fixing the stellar temperature to this value for the remainder of the analysis in the paper. 

This stellar temperature is about 350\,K above the estimated value from the broad band colours. The difference could be due to spots and variability not being included in the previous estimates. We will use the stellar temperature determined from our light curves as a fixed value for the remainder of our analysis. Small changes of up to a few 100\,K in this value do not introduce significant changes in the deduced spot properties. In particular, variations in the spot properties over time are robust against changes in the adopted stellar temperature.

As a side note, the same least square optimisation fit has been attempted with only a subset of the available light curves. In particular when using only $V$, $R_c$ and $I_c$ amplitudes, we do not find any reliable results for the stellar temperature, and indeed at most epochs the procedure does not converge at all. The reason is that with this subset of filters, only one side of the spectral energy distribution is sampled and thus the stellar temperature cannot be deduced.

\subsection{Spot properties during high cadence observations}\label{hc_analysis}

The high cadence data (between MJD\,=\,58330 to 58400) allow us to verify how reliable the spot properties (temperature and coverage) can be determined from amplitudes in only a subset of the filters, when the stellar temperature is fixed to the above determined 5050\,K. This is important as for the vast majority of the \vc\ light curve only $V$, $R_c$ and $I_c$ amplitudes are available.

\subsubsection{Least square optimisation}

\begin{figure*}
\centering
\includegraphics[width=\columnwidth]{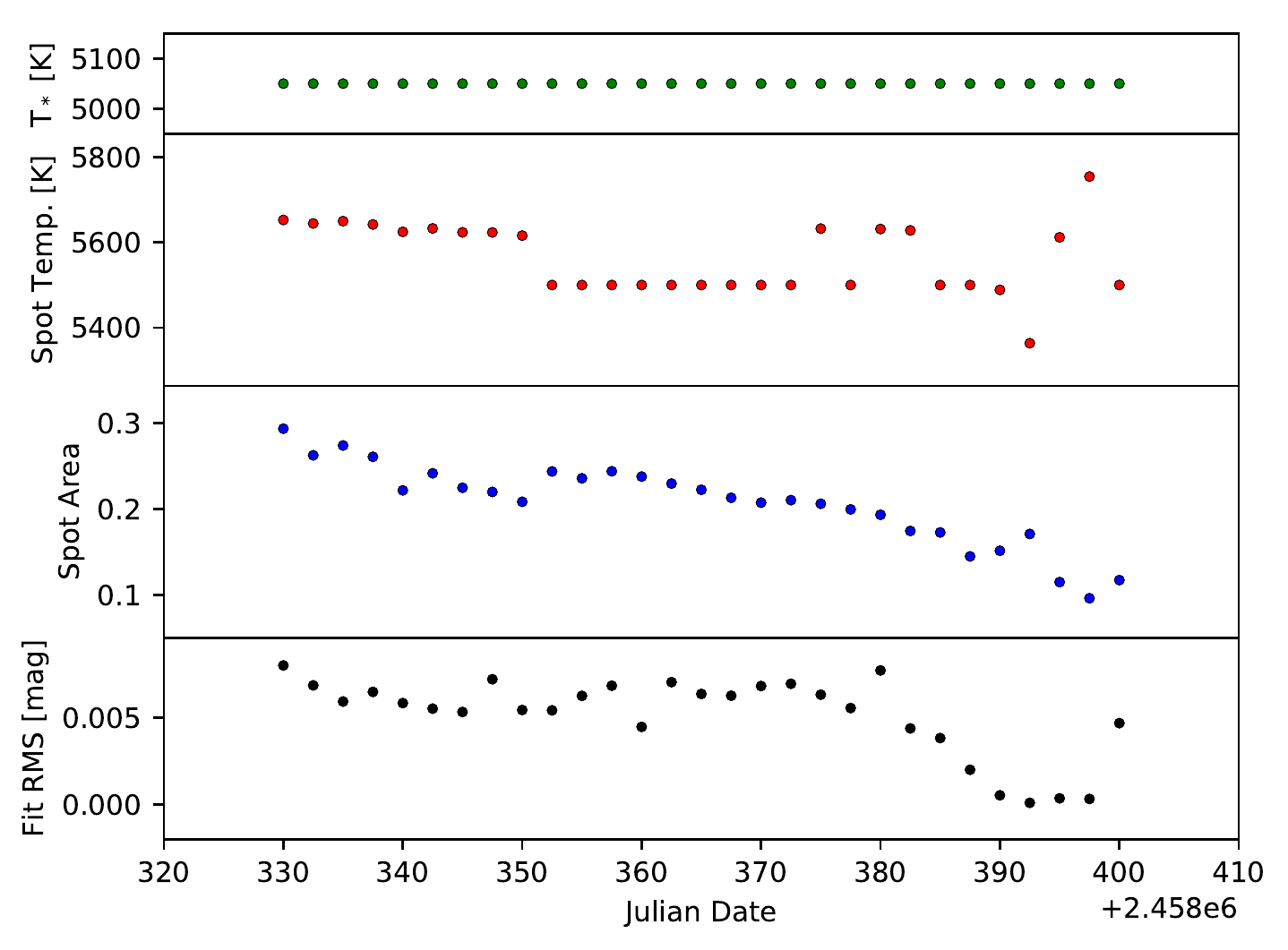}\hfill
\includegraphics[width=\columnwidth]{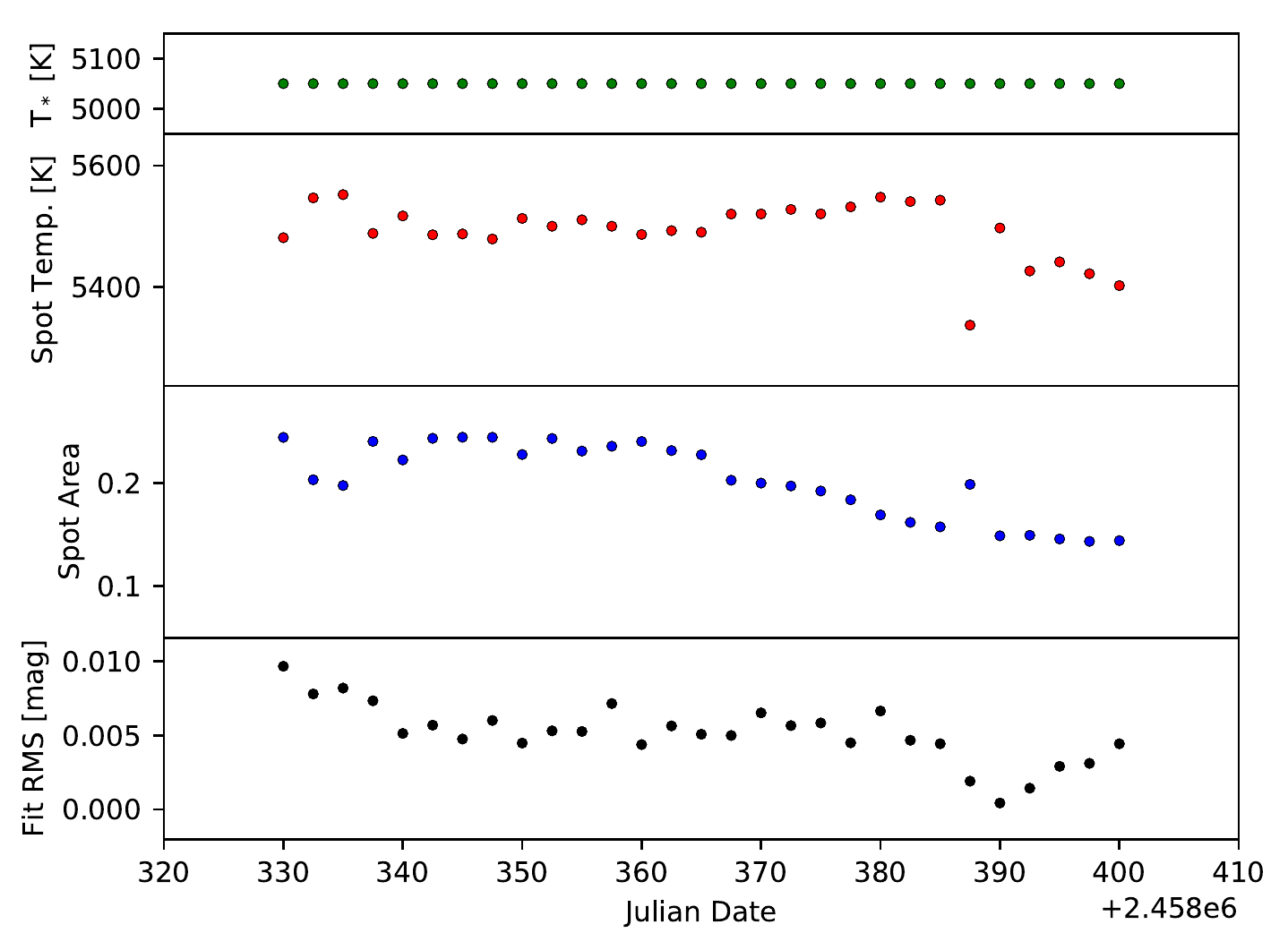} \\
\includegraphics[width=\columnwidth]{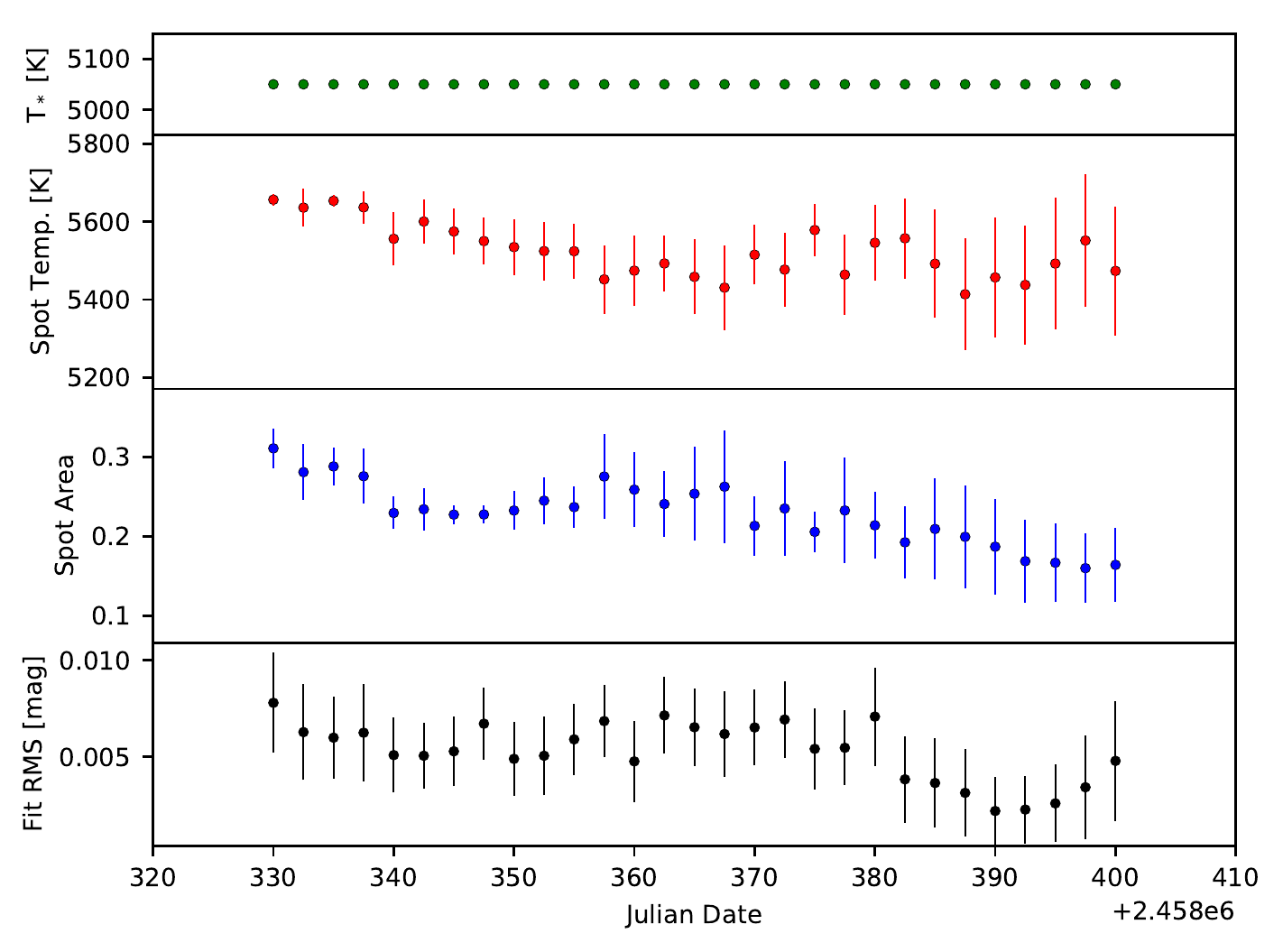}\hfill
\includegraphics[width=\columnwidth]{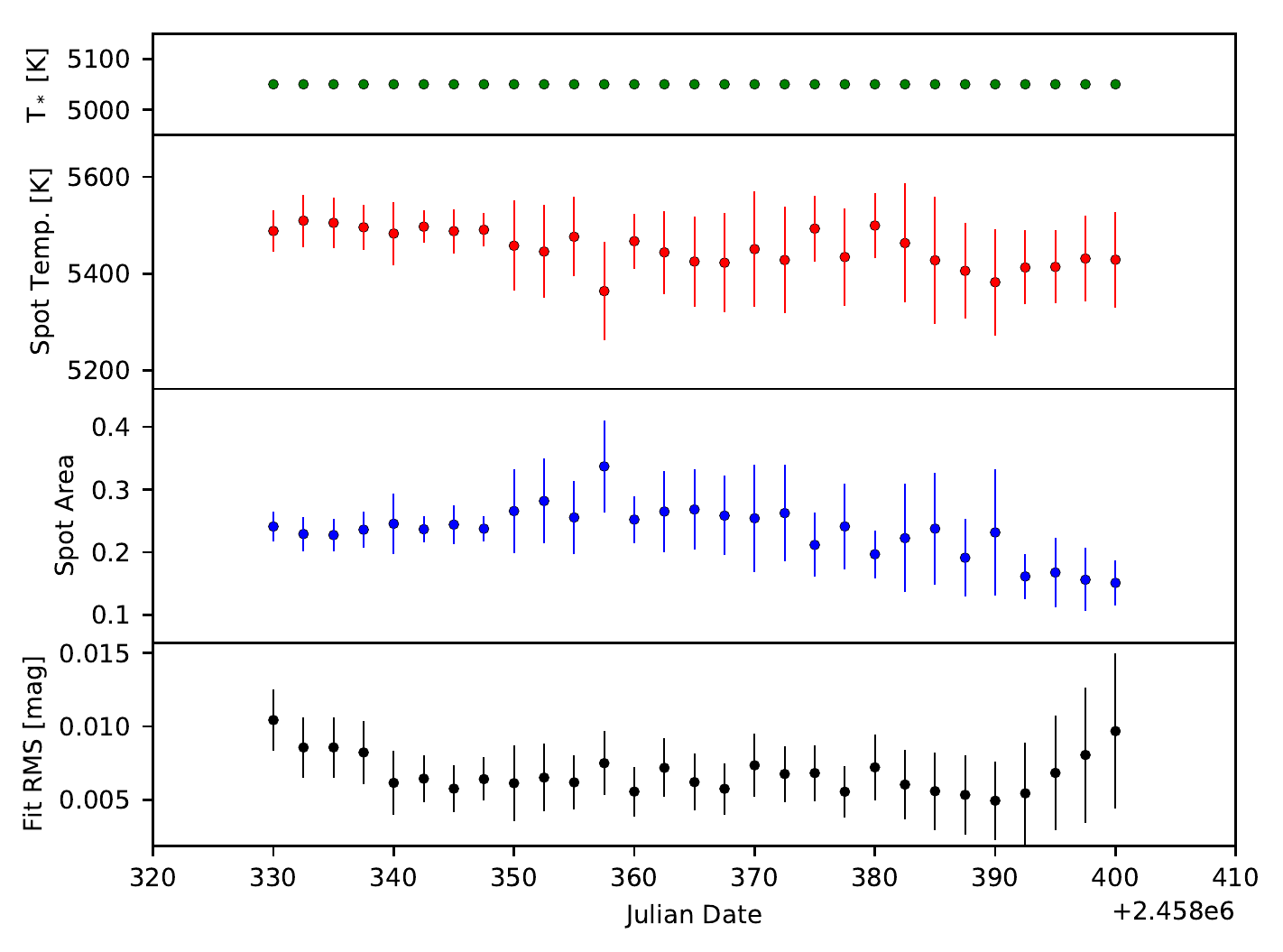} \\
\caption{\label{fig_test4} Stellar temperature, spot temperature, spot coverage and $rms$ of the fit for the high cadence data. In the left column only $V$, $R_c$ and $I_c$ data has been used, while in the right column all broad band filters have been included. In the top row the least square optimisation method has been utilised, while in the bottom row we have used a model grid. In all cases a fixed stellar temperature of 5050\,K was used.}
\end{figure*}

We now repeat the least square optimisation procedure discussed above to obtain the spot properties, but with fixed stellar temperature. Since we know from using all filters (see Sect.\,\ref{findtstar} and Fig.\,\ref{fig_test5}) that the spots are warmer than the star, we select as initial guesses for the spot temperature values between 5100 and 6100\,K and for the spot coverage values between 0.05 and 0.45. The fit is repeated 50 times with the initial start values chosen randomly from a homogeneous distribution within the above set ranges. The best spot parameters are selected from the fit with the lowest $rms$ in the 50 repeats. We show the results of this procedure in the top panels in Fig.\,\ref{fig_test4}. In the right column, all broad band filters are used for the fit. For the left column we use only $V$, $R_c$ and $I_c$.

When using all broad band filters, we find that the spot temperature varies between 5400 and 5600\,K, while the spot coverage is between 0.15 and 0.25. When removing \au\ and \ab\ from the fit, values outside these ranges are found in some cases. In particular unphysical spot temperatures jumps are occasionally found. These jumps are not considered real, considering that the light curve amplitudes and thus the spot properties are averaged over a $\pm$15\,d window. In other words any changes in spot properties should be happening at this time scale or slower.

When excluding the spot temperature outliers and jumps, the results obtained from different sets of filters for the spot temperature and coverage are in principle comparable. The $rms$ of the fits (bottom row in all plots in Fig.\,\ref{fig_test4}) is very similar for all cases. Within the time period analysed here, the spot coverage decreases by about 50\,\%, while the spot temperature varies by about 100\,K. 

\subsubsection{Model amplitude grid}

To further investigate the spot properties and to find the cause for the unrealistic jumps, we use a grid-based approach, as opposed to the least-squares optimisation. We calculate a grid of model amplitudes for all broad band filters with our fixed stellar temperature. In the grid the spot temperature ranges from 5100\,K to 6000\,K in steps of 5\,K and the spot coverage ranges from 0 to 0.4 in steps of 0.0025. For each set of broad band amplitudes the $rms$ of the model amplitudes from the measured amplitudes is determined. The spot properties resulting in the lowest $rms$ are selected as the best fit. 

Similar to estimating the uncertainties of the light curve amplitude, we estimate the uncertainties of the spot properties by repeating the fit $N=10000$ times. Each time we add a random number to the broad band amplitudes, where the random number is drawn from a Gaussian distribution with a mean of zero and a sigma equal to the uncertainty of the amplitude. The mean and standard deviation of the 10000 individual best fit spot parameters from this procedure are then selected as best fit spot temperature and coverage and their uncertainties.  

\begin{figure*}
\centering
\includegraphics[width=\columnwidth]{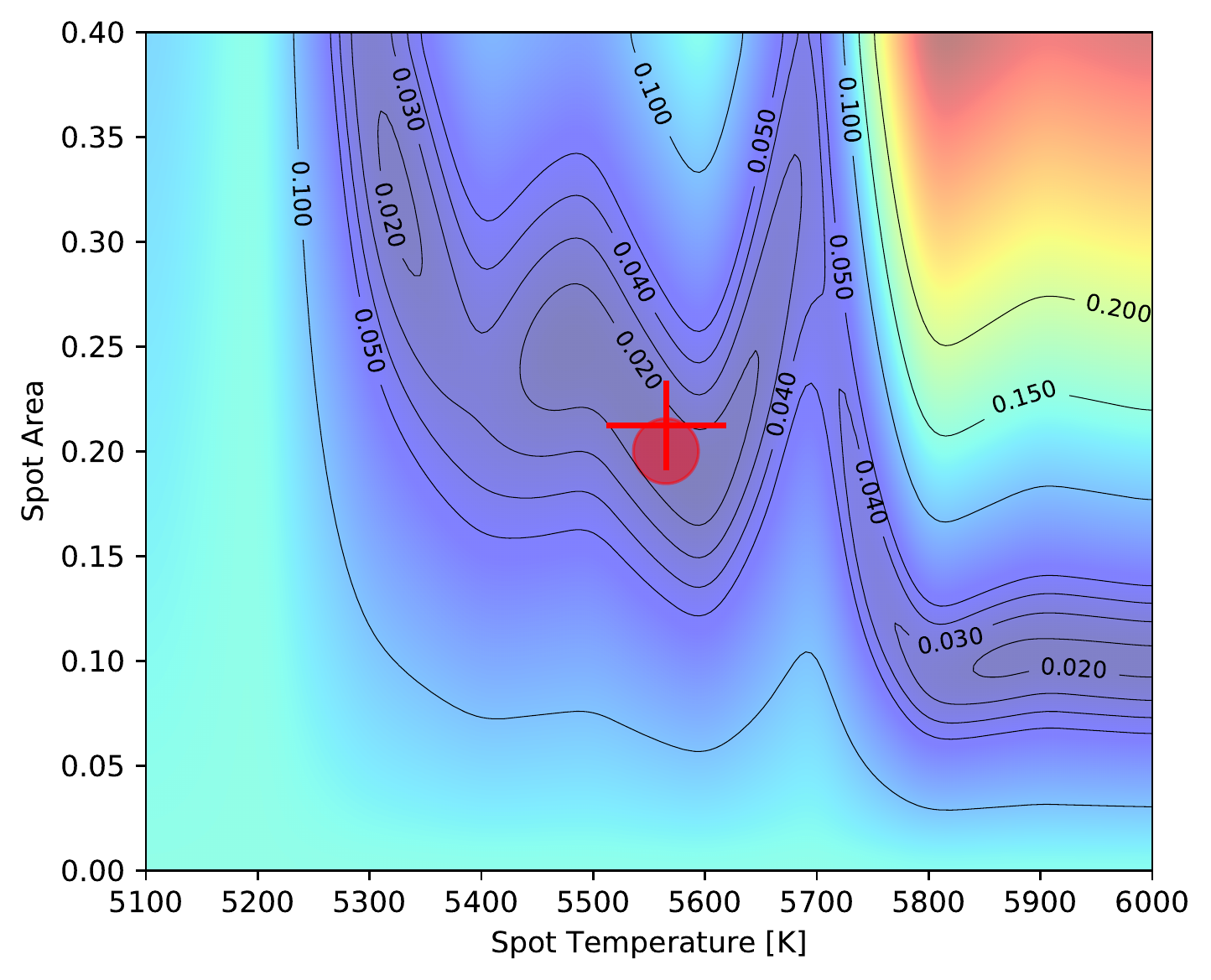} \hfill
\includegraphics[width=\columnwidth]{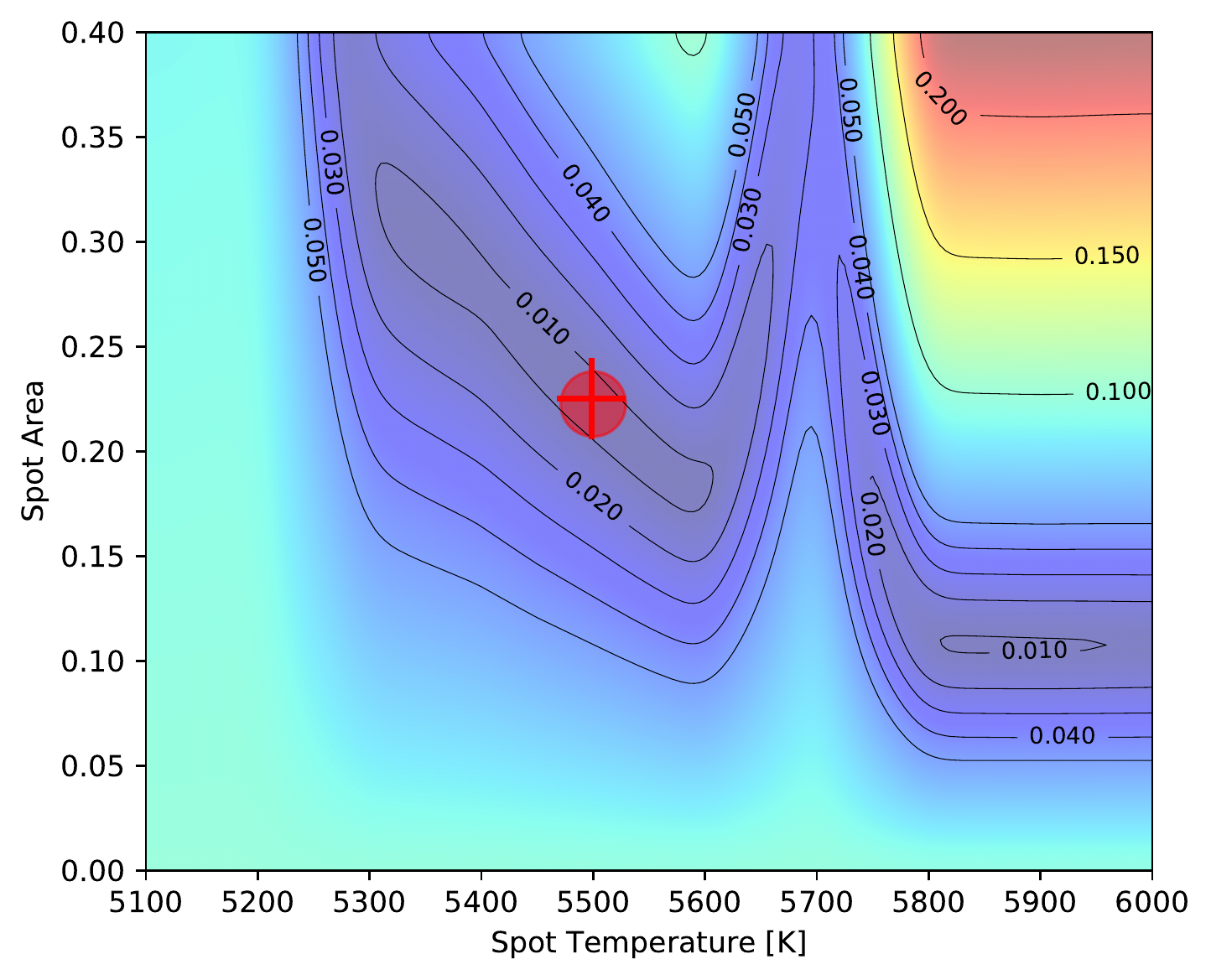}
\caption{\label{fig_spotcont} Smoothed $rms$ contours as a function of spot temperature and coverage, for a specific set of amplitudes. The $rms$ in magnitudes is indicated. In the right panel, all filters are used, whereas the left panel shows results only for $V$, $R_c$, and $I_c$. The circle indicates the actual minimum position in this particular $rms$ plot, while the error cross gives the mean and standard deviation for the lowest $rms$ position in all 10000 iterations.}
\end{figure*}

In Fig.\,\ref{fig_spotcont} we show in colour code and contours the $rms$ of one randomly selected set of amplitudes as a function of spot properties. On the left only $V$, $R_c$ and $I_c$ data is used, while on the right all broad band filters are included. The overlaid cross indicates the position of the average best fitting spot temperature and coverage. The size of the cross indicates the uncertainties of both values estimated from all 10000 iterations The circle marks the position with the minimum $rms$ for the amplitudes used to generate this particular plot.

For the example in Fig.\,\ref{fig_spotcont}, a spot coverage of 0.20 and a spot temperature of 5600\,K is estimated, well within the range derived using the least-square optimisation for the whole high-cadence data set. A detailed look at Fig.\,\ref{fig_spotcont} also explains the abrupt changes in the parameters found with the least square optimisation. Several minima in the $rms$ are found across the range of parameters, some of them with quite similar $rms$. Thus, small changes in the amplitudes can strongly affect the outcome. Also, a spot temperature of 5700\,K clearly results in bad fits. When all filters are available, the minimum has a much better defined location. When only $V$, $R_c$ and $I_c$ are used, more local minima appear which are not as well defined. Thus, for $V$, $R_c$ and $I_c$ only light curves, the spot parameters are generally less well determined for this particular object.

In Fig.\,\ref{fig_test4} we compare the spot properties obtained using the model grid (bottom row) with the least square optimisation procedure (top row), using the same filter combinations. As can be seen again, the general behaviour for the spot parameters obtained with the model grid is similar for all filter combinations, and it also agrees with the least square optimisation results within the uncertainties estimated from the model grid. During the period when we have high cadence data, the spot temperature has remained mostly constant, while the coverage of the spot has dropped by about 50\,\% from 0.3 to 0.15.

Comparing the results from the model grid for $V$, $R_c$ and $I_c$ with those from all broad band filters shows that there are no large or systematic differences between the spot properties (bottom row in Fig.\,\ref{fig_test4}). This gives us re-assurance that the results for the entire data set (when mostly only $V$, $R_c$ and $I_c$ is available) will provide robust results when the model grid method is used, contrary to the nonphysical jumps found when using the least square optimisation method (see top row in Fig.\,\ref{fig_test4}). The model grid method is also computationally by far faster than the least square method and reliable one sigma uncertainties can be readily obtained for all inferred spot parameters. Therefore, in the next Sect.\,\ref{longterm}, we will analyse the long-term light curve only using the model grid method.

\subsection{Amplitude and spot properties for the entire light curve}\label{longterm}

\begin{figure}
\centering
\includegraphics[width=\columnwidth]{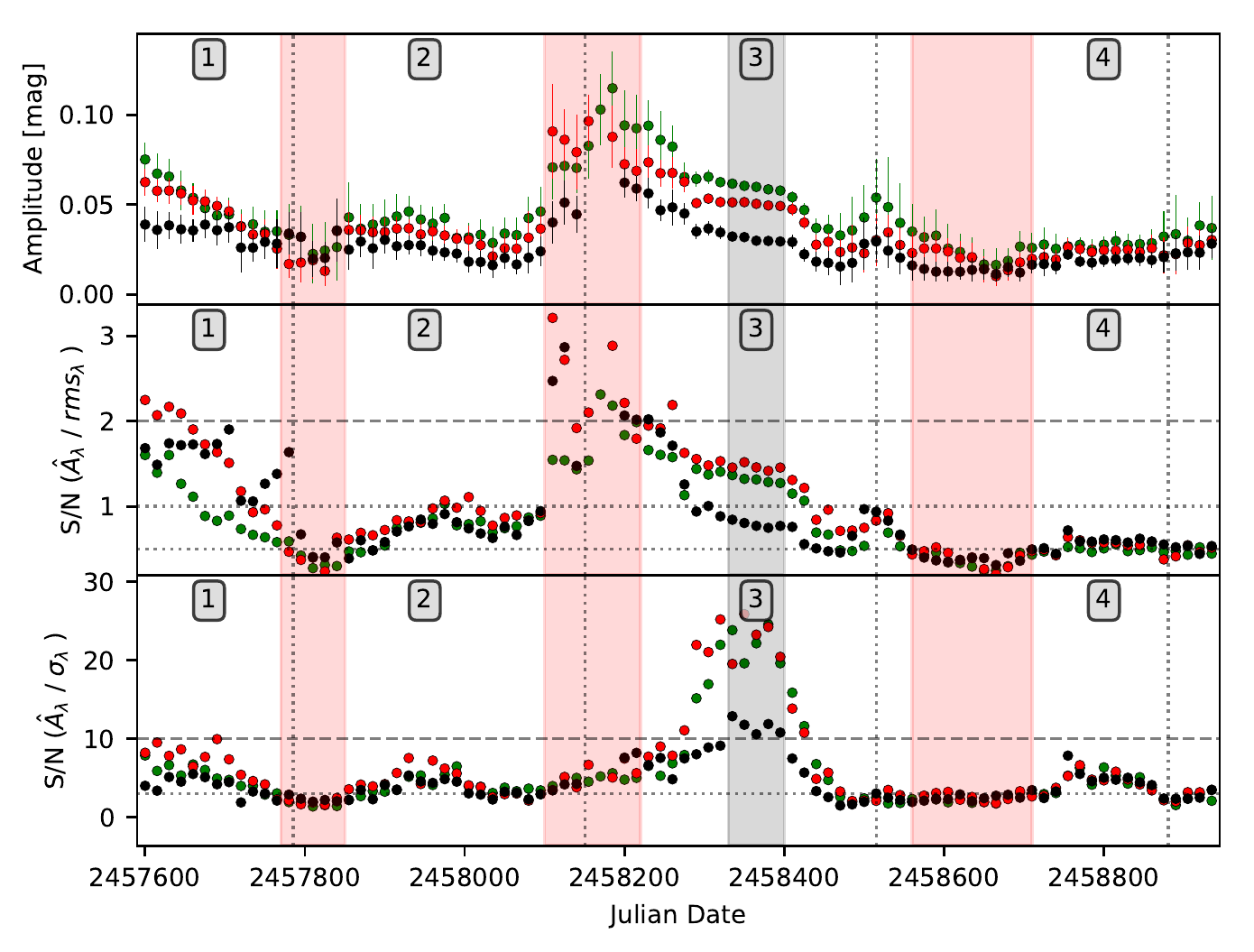}\\
\includegraphics[width=\columnwidth]{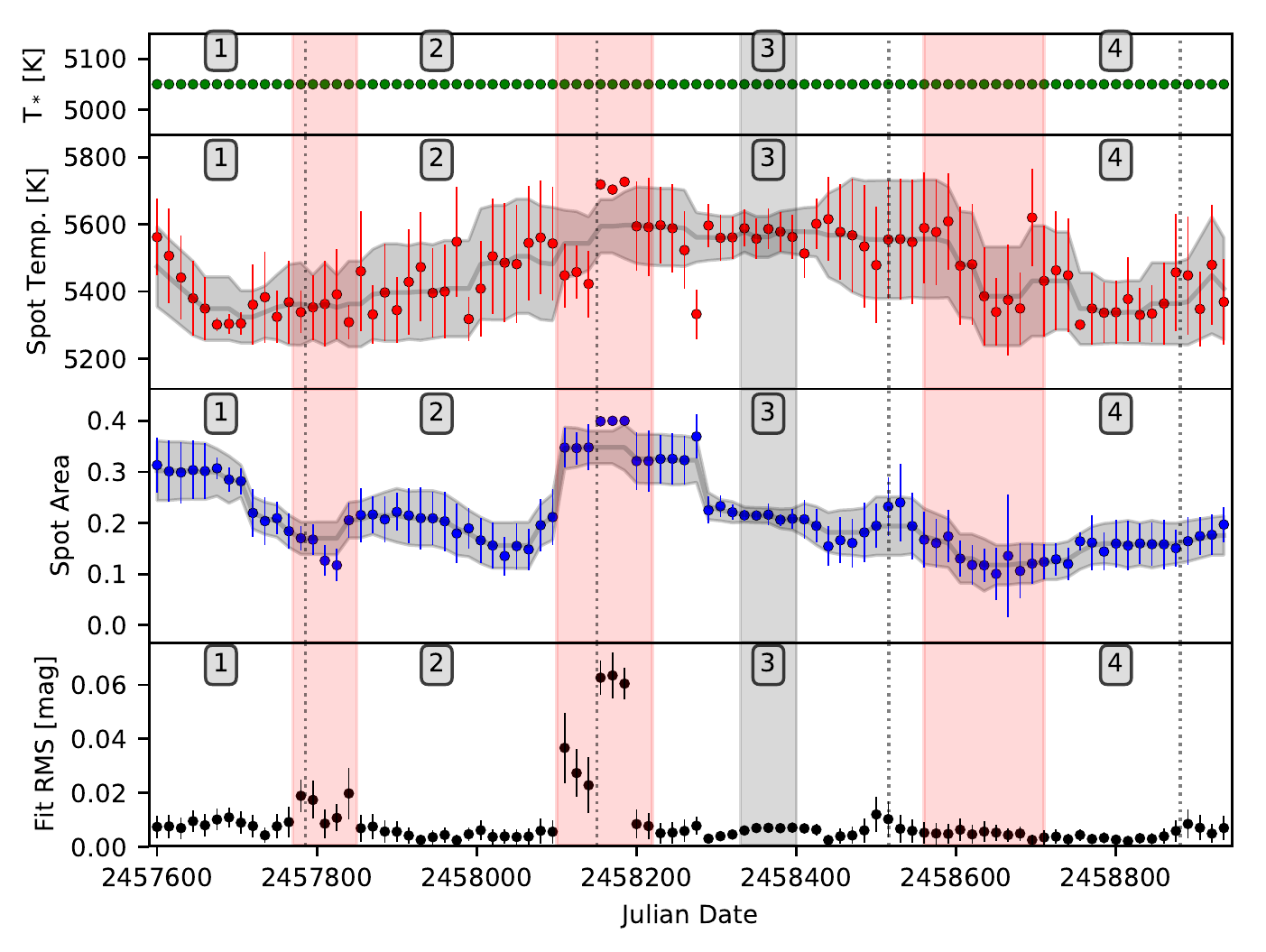}
\caption{\label{fig_test6} Long term \vc\ amplitude (top) and spot (bottom) properties using $V$, $R_c$ and $I_c$ data only. All values are determined over a $\pm$\,60\,d window. The same properties as in Fig.\,\ref{fig_test5} are shown. We have overlaid in gray the running median and uncertainty for the spot temperature and coverage. Red hashed areas indicate times with low signal to noise or rapidly changing spot properties, which have been excluded from our analysis. The dotted vertical lines indicate the worst visibility, at Feb 1st each year. The gray hashed area indicates the time with the high cadence data. The numbers one to four indicate the phases of the light curve that are discussed in detail in Sect.\,\ref{evolution}.}
\end{figure}

All analysis for long term trends is based on the $V$, $R_c$ and $I_c$ light curves, which are available for the entire data set. We have chosen to determine light curve amplitudes in a $\pm$60\,d window, given the typical cadence of the \hc\ data. This window results in only a very small number of epochs where the amplitude fit fails. Note that changing the exact value of the time window width has no influence on the results. We perform the fit at 90 epochs, equally distributed over the entire 4\,yr baseline, i.e. every 15\,d.

The results of this procedure are shown in Fig.\,\ref{fig_test6}. The upper part of the figure relates to the light curve parameters. In the top panel we display the amplitudes and their uncertainties for the three used broad band filters. The second panels shows the ratio of the amplitudes and the $rms$ of the photometric data after subtraction of the fitted sine function, i.e. the typical photometric uncertainty of the individual brightness measurements. The bottom panel shows the signal to noise ratio of the amplitudes itself. In all cases the amplitude values are the mean of the individual measurements from the 30 realisations of the individual photometric data considering their uncertainties (as discussed in Sect.\,\ref{ampfitting}). The uncertainties represent the $rms$ from this mean.

The bottom part of Fig.\,\ref{fig_test6} shows the estimated spot properties and their uncertainties. In the 1st panel the stellar temperature is displayed, which is fixed to 5050\,K. In the second and third panel we plot the spot temperature and the spot coverage with uncertainties. We further over plot a running median and a running average uncertainty range in both panels. For the interpretation of the results, we use these running median averages, as they ensure that potentially remaining nonphysical outliers are removed. The forth panel shows the $rms$ of the fit, i.e. the $rms$ of the measured from the fitted amplitudes. 

In all panels of Fig.\,\ref{fig_test6} the vertical dotted lines indicate February 1st each year, which is the time of worst visibility of the object. The grey shaded area from times JD\,=\,2458330 to 2458400 represents the high cadence part of the light curve which has been investigated in detail in Sect.\,\ref{hc_analysis}. Furthermore, the red shaded areas represent periods when the signal to noise ratio of the determined amplitudes is low, or where the spot properties seem to be changing much more rapidly than the time window over which the amplitudes are determined. The resulting spot properties are discussed in detail in Sect.\,\ref{discussion}.

\section{Discussion}\label{discussion}

\subsection{Review of methodology}

Our method of fitting light curve properties, as discussed in Sect.\,\ref{ampfitting}, is able to reliably determine the amplitude of light curves (or parts of them) for \hc\ targets, provided an initial guess for the period is available from a periodogram analysis. We show here that for a typical \hc\ coverage we are able to determine amplitudes with a signal to noise of three, that are as low as half of the typical uncertainty of the individual photometric data points. When more data points are available, the accuracy improves further. Note that the accuracy also depends on the actual period of the object, and should improve for longer periods. The example of the very fast rotator \vc\ presented here shows that the method is readily applicable to most other stars in the \hc\ data set with typical YSO rotation periods. The only adjustment to be made is the time window over which the light curve properties are to be fit.

Utilising the available contemporaneous, multi-filter amplitudes combined with model stellar atmospheres, we are able to estimate stellar and spot properties. Generally, \hc\ data in all optical broad band filters is most likely needed to evaluate the stellar temperature. However, it might be possible for certain effective stellar temperatures to obtain reliable results without the $I_c$ or $U$ data. With a fixed or known stellar temperature, and using a model grid approach with subsequent median filtering, we find that for a typical K-type star the usage of $V$, $R_c$ and $I_c$ data is sufficient to reliably estimate spot temperature and coverage. This approach reproduces the measured light curve amplitudes to better than 0.01\,mag (see bottom panel in Fig.\,\ref{fig_test6}). Typically, the spot temperature can be estimated with an uncertainty of about 100\,K, while the fractional coverage of the spot is accurate to within 3\,--\,4\,\%. These uncertainties will depend slightly on the specific stellar surface temperature, observing cadence and available filters. In particular, in some cases several combinations of spot temperature and spot coverage can lead to similarly good fits. To capture and evaluate such cases, we suggest to follow our approach of determining amplitudes in an over-sampled fashion along the light curve. The resulting light curve and spot property evolution can then be more reliably traced and interpreted from the running median of the fit results.

\subsection{Spot properties on \vc}

The spot modelling over the entire light curve gives us a four year baseline to examine the spot properties of this particular star. Spot activity is visible in the entire period, with notable variability on timescales from several months to years. In this subsection we focus on the overall spot properties, whereas in the next subsection we look at the spot evolution.

The spot temperatures range from 5300 to 5700\,K in our observing period, with typical error bars of $\pm 100$\,K (see Fig.\,\ref{fig_test6}, lower panel). That means, the spots are always warmer than the rest of the photosphere. At no time during the observing period do we find that spot temperatures below the stellar temperature lead to the best fit. The spot coverage is in the interval between 15\,\% and 40\,\%, although given that the signal-to-noise ratio for the amplitude drops to marginal levels in parts of the light curve (see Fig.\,\ref{fig_test6}, upper panel), the 15\,\% should be considered an upper limit on the minimum spot coverage. 

As explained in Sect. \ref{target}, \vc\ is either a close (but detached, non-eclipsing) active binary (a RS CVn star), and/or a pre-main sequence star without disk accretion (a weak-line T Tauri or WTTS). A broad overview of spot properties on various types of active objects and from various methods is given in \citet{2005LRSP....2....8B} and \citet{2009A&ARv..17..251S}. Both WTTS and RS CVn-types can have enormously large spots covering half or more of a hemisphere. Our values for \vc\ are certainly within the plausible range for these kind of stars, in particular when considering that \vc\ is a very fast rotator driving strong magnetic activity. Thus, in terms of spot coverage, \vc\ fits into the general phenomenology observed for other stars of this nature. We also note that light curve amplitudes can only measure lower limits on spot coverage, since they are only sensitive to the asymmetrically distributed portion of spots on the stellar surface.

For spot temperatures, however, the consensus is that cool spots dominate on active stars. One explanation for hot or warm spots is accretion, either in form of disk accretion in young stars \citep{1993A&AS..101..485B} or as mass transfer in semi-detached or contact binaries (see summary by \citet{2019PASJ...71...21K} and references therein). Our target does not show any evidence for disk accretion or mass transfer, hence these options can be ruled out. 

 Without indications of accretion, it is plausible that the warm spots are caused by magnetic activity, i.e. the 'spots' are not dark regions in the photosphere, analogous to Sun spots, but constitute warmer areas higher in the atmosphere, comparable to chromospheric plage. Plage is a common feature in RS CVn-type stars \citep[e.g.][]{2014NewA...32....1Z}, with plage and spots being closely spatially related \citep{2008A&A...479..557F}. There is good evidence that (low-activity) main-sequence stars get brighter with increased activity, because plage dominates over spots. For pre-main sequence stars the opposite appears to be the case \citep{2004AAS...204.0304L,2007ApJS..171..260L}. Thus, the warm spots seen in our analysis may indicate that the object is in fact a binary, and not in the pre-main sequence stage. However, see \citet{2016MNRAS.463.4383K} for a discussion of the possibility of above-photospheric spot temperatures in WTTS. Furthermore, the surface brightness reconstructions made with Zeeman Doppler imaging for WTTSs by \citet{2014MNRAS.444.3220D, 2015MNRAS.453.3706D, 2017MNRAS.465.3343D} and \citet{2017MNRAS.472.1716H, 2019MNRAS.484.5810H} indicate the presence of cool spots and warm plages at the stellar surface of young stars with convective envelopes. Thus, the warm spots found for \vc\ do not provide conclusive evidence for the binary nature of the source.
 
\subsection{Spot evolution}\label{evolution}

As evident from Fig.\,\ref{fig_test6} there are clear variations in variability amplitudes, and the modelled underlying spot properties. The spot temperatures have quite large error bars, hence trends are difficult to exactly determine. On the other hand, the spot coverage has small uncertainties, with variations larger than the errors. 

We have identified four distinct phases of spot activity during the four years of \hc\ observations. They are separated by periods of very little change (JD\,=\,2457800 and 2458650) or very rapid change (JD\,=\,2458150) in variability amplitude. During these phases the light curve is presumably governed by one dominant spot or group of spots, resulting in the sine-curve like folded light curves seen e.g. in Fig.\,\ref{lombscarcle_V}.

\begin{figure}
\centering
\includegraphics[width=\columnwidth]{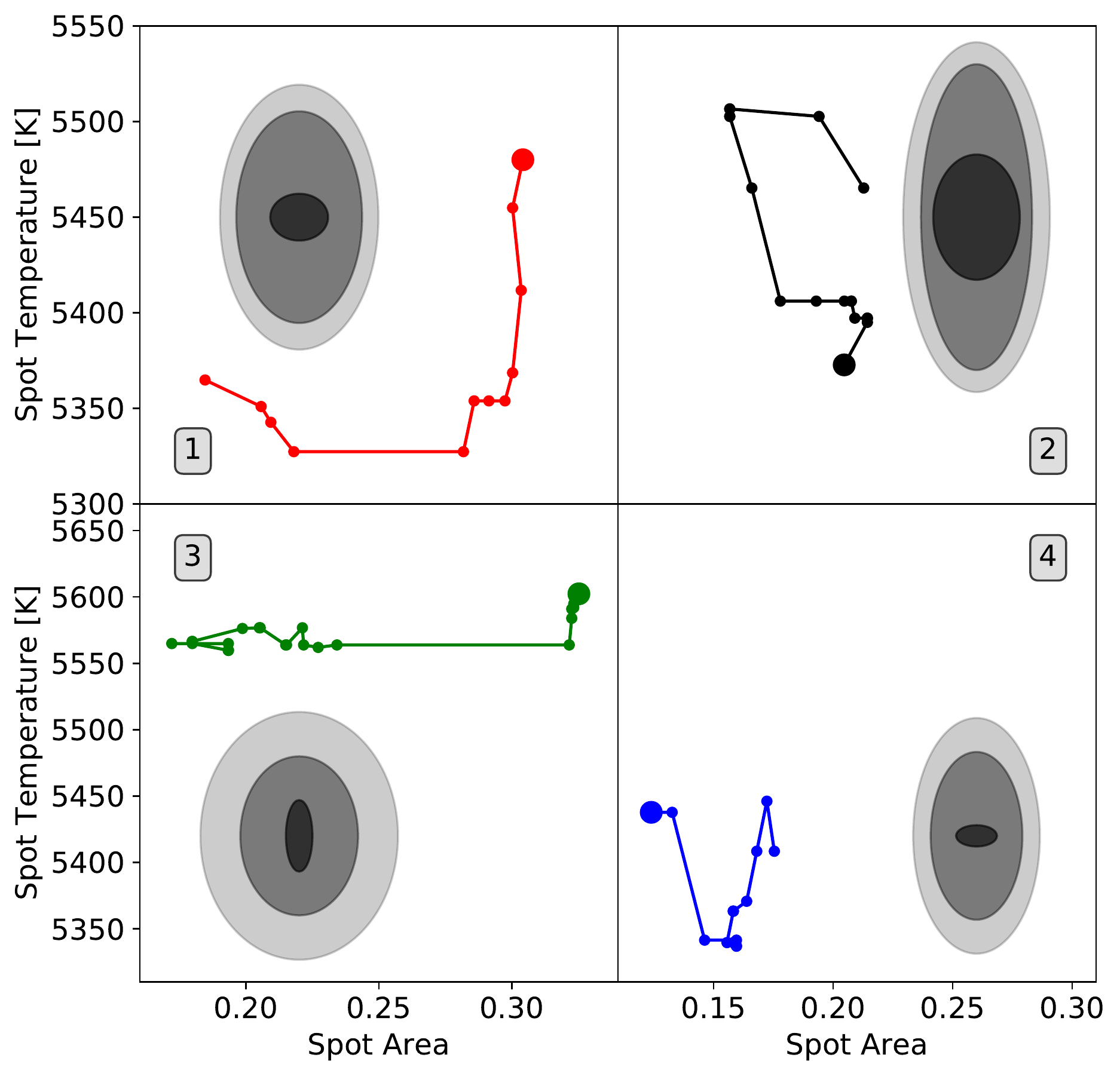}\\
\caption{\label{fig_spottempsize} Spot temperature vs. spot coverage evolution for the four phases discussed. Shown are the running median values as in Fig.\,\ref{fig_test6}. The numbers in each panel indicate the phases shown in Fig.\,\ref{fig_test6} and discussed in Sect.\,\ref{evolution}. The dark gray ellipses show the maximum (light gray), median (gray) and minimum (dark gray) uncertainties of the data points in each panel. The largest data point in each panel is the first one in time, and there is a 15\,d gap between successive data points.}
\end{figure}

We have indicated the different phases by numbers one to four in Fig.\,\ref{fig_test6} for clarity. The periods in between phases are highlighted as red shaded areas. Below we give a very brief description of the inferred spot behaviour during the four phases. The spot property evolution for each phase can also be seen in more detail in Fig.\,\ref{fig_spottempsize}, where we show the running median spot properties and the typical uncertainties in spot coverage vs. spot temperature plots. 

\begin{itemize}
    
\item {\bf Phase\,1:} A general decline of amplitudes in $V$ and $R_c$, while \ai\ remains almost constant. During the first half, the spot temperature declines from about 5500\,K to 5300\,K at constant spot coverage (0.3), followed by a shrinking of the spot coverage to 0.2 at constant temperature. The initial decrease of temperature occurs at a  rate of about $-$2.2\,K/d. Overall, this is a phase where spot activity declines.

\item {\bf Phase\,2:} There are only minor variations in the amplitudes, with a maximum in \av\ roughly in the middle of the phase. The spot temperatures are constant (5450\,K) within the uncertainties, with a weak trend of slightly increasing values at a rate of about $+$0.2\,K/d. The spot coverage (0.2) is also constant within the uncertainty. 

\item {\bf Phase\,3:} There is a clear trend of decreasing amplitudes in all filters, by at least a factor of two. As for the previous phase the spot temperatures do remain constant within the uncertainties at about 5600\,K. The spot coverage decreases from 0.35 to 0.2, with a small increase towards the end of this phase. This phase contains the high cadence data (indicated as gray in Fig.\,\ref{fig_test6}). Due to the large amount of data the amplitude uncertainties are much lower, and hence also the spot temperature and coverage uncertainties. The analysis of this data with a much smaller $\pm$15\,d window is shown in the bottom left panel of Fig.\,\ref{fig_test4}. The results are consistent with the results obtained for the entire light curve. They show the spot temperature decreasing from 5650\,K to 5500\,K. At the same time the spot coverage drops from 0.35 to 0.2. Overall, the characteristics of phase 3 are similar to phase 1, but at slightly higher spot temperatures.

\item {\bf Phase\,4:} The amplitudes remain constant for a large part, but start to slowly increase towards the end. Accordingly the spot temperature (5400\,K) and coverage (0.15) remain constant throughout, and only increase towards the end. The temperature increases at a rate of $+$2.9\,K/d towards the end. 

\end{itemize}

We note that the duration of all phases is significantly longer than our time window of $\pm$60\,d, and they are hence unlikely to be an artefact. One possibility to interpret these phases is as lifetime of large spots or groups of spots. If that is the correct interpretation, the spot lifetime in the four phases would be $>$180\,d, 230\,d, 360\,d, $>$230\,d, overall, or between half a year and one year. These values are consistent with spot lifetimes determined in the literature for active stars. For example, \citet{2017MNRAS.472.1618G} used Kepler data to estimate spot decay timescales between 20 and 300\,d for K-stars. A fast rotating single star like \vc\ would be expected to have a shorter spot lifetime, but a potential close companion and tidal coupling may extend the lifetime. 

As described in Sect. \ref{general}, we find a long-term change in the photometric brightness level, when combining our lightcurve with archival data. The approximate period of this change is in the range of 6\,--\,6.5\,yr. The most straightforward interpretation is the presence of a magnetic activity cycle, equivalent to the 11\,yr cycle in the Sun. The duration of $\sim$\,6\,yr seems reasonable when comparing with typical cycle periods for active fast rotating K stars \citep{2009A&A...501..703O} and active binaries \citep{1995ApJS...97..513H}.
 
\section{Conclusions}

We have utilised the wealth of photometric data available from our \hc\ citizen science project \citep{2018MNRAS.478.5091F} to investigate the properties of the active star \vc\ over a period of four years. As demonstrated in \citet{2020MNRAS.493..184E}, the internal calibration of the \hc\ data allows us to correct colour terms in the photometry caused by the variety of different optical filters used by the participating members in the project. This optimises the usefulness of the homogeneous data set for any kind of study relying on long term photometric monitoring. We can achieve median photometric uncertainties in our data of the order of a few percent.

In this paper, we have developed a robust, reliable and automated procedure to determine amplitudes of photometric variability in our light curves, including an accurate estimate of the statistical uncertainties. Applying these methods to the \hc\ data set of \vc, we find that this source is periodically variable in all bands, and a very fast rotator with a period of 0.8246\,$\pm$\,0.0004\,d. The amplitudes increase with decreasing wavelength, and are time variable. Our methodology allows us to determine amplitudes with a signal to noise of three or better, which are as low as half the typical photometric uncertainty of the individual photometric data points in the light curve. The \hc\ cadence of data has enabled us to determine the light curve properties for multiple epochs within $\pm$60\,d windows, i.e. we are able to investigate changes in behaviour of \vc\ on time scales of a few months.

We have used the PHOENIX model stellar spectra \citep{2013A&A...553A...6H}  to model the spot temperature and coverage that can simultaneously explain the measured amplitudes in the various optical filters available for \vc\ from \hc. We find that a grid of models, combined with error propagation and temporal oversampling, plus a final median filtering allows us to accurately, robustly and automatically trace the spot evolution on the object. The spot temperature can typically be estimated to within 100\,K, while the spot coverage has an uncertainty of 3\,--\,4\,\%. In \vc\ the spots are typically a few 100\,K warmer than the stellar surface and cover between 10\,\% and 30\,\% of the visible hemisphere, plausibly caused by chromospheric plage. The spots evolve on timescales of six to twelve months, which can be interpreted as the lifetime of the spots or group of spots on the stellar surface. 

Based on auxiliary photometric data from other public surveys and data sets, with a total baseline of 11\,yr, we find a general 6\,yr long term photometric cycle for \vc, which we interpret as a general activity cycle of the star. The position of the object in the Gaia HR diagram suggests either a roughly equal mass, close (detached, non-eclipsing) binary - a RS\,CVn star -  or a fast rotating, more than 20\,Myr old pre-main sequence weak-line T\,Tauri star. Our currently available data does not allow us to definitely decide between the two possible interpretations about the nature of the source. A short sequence of high resolution spectroscopy, covering one or two periods, should be sufficient to establish if the source is indeed a binary.

\section*{Acknowledgements}

We would like to thank all contributors of observational data for their efforts towards the success of the \hc\ project.
A.\,Scholz acknowledges support through STFC grant ST/R000824/1. 
We acknowledge with thanks the variable star observations from the AAVSO International Database contributed by observers worldwide and used in this research. This research made use of Astropy,\footnote{\tt http://www.astropy.org} a community-developed core Python package for Astronomy \citep{2018AJ....156..123A, 2013A&A...558A..33A}.

\section*{Data Availability Statement}

The data underlying this article are available in the HOYS database at http://astro.kent.ac.uk/HOYS-CAPS/.


\bibliographystyle{mnras}
\bibliography{bibliography} 


\begin{figure*}
\appendix
\section{Lomb-Scargle Diagrams}\label{lombscarcle_all}
\centering
\includegraphics[width=0.99\columnwidth]{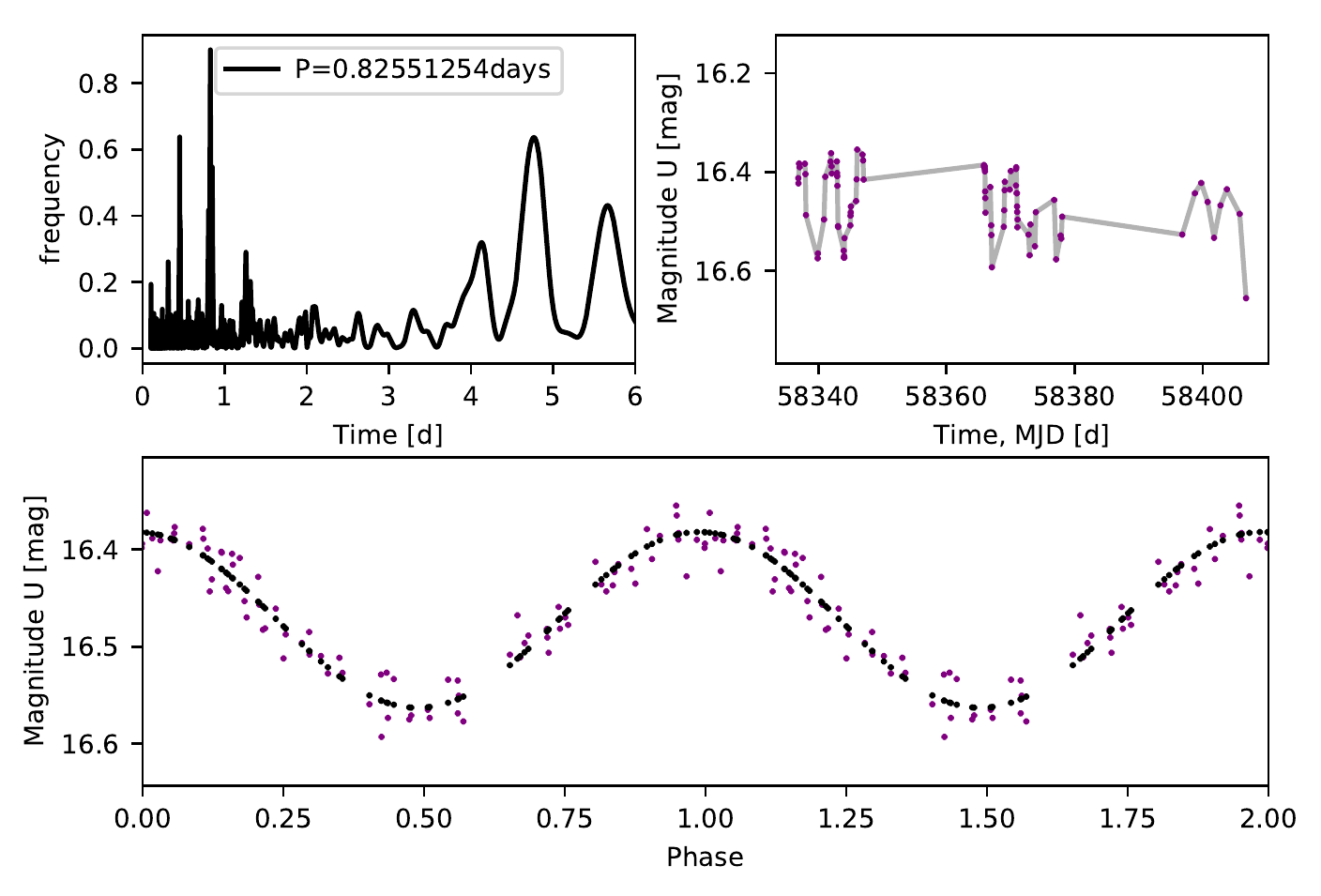} \hfill
\includegraphics[width=0.99\columnwidth]{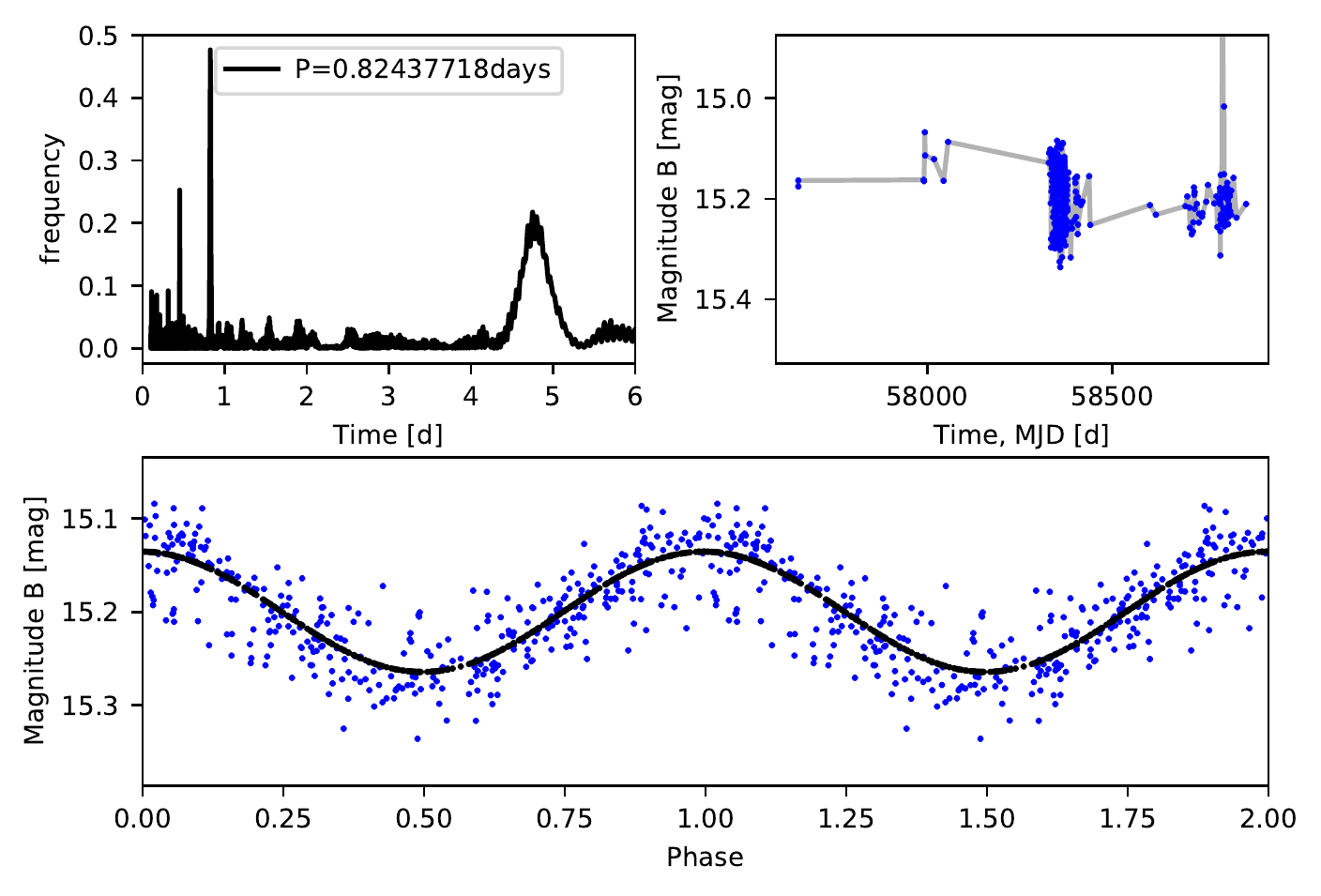} \\
\includegraphics[width=0.99\columnwidth]{test_V.pdf} \hfill
\includegraphics[width=0.99\columnwidth]{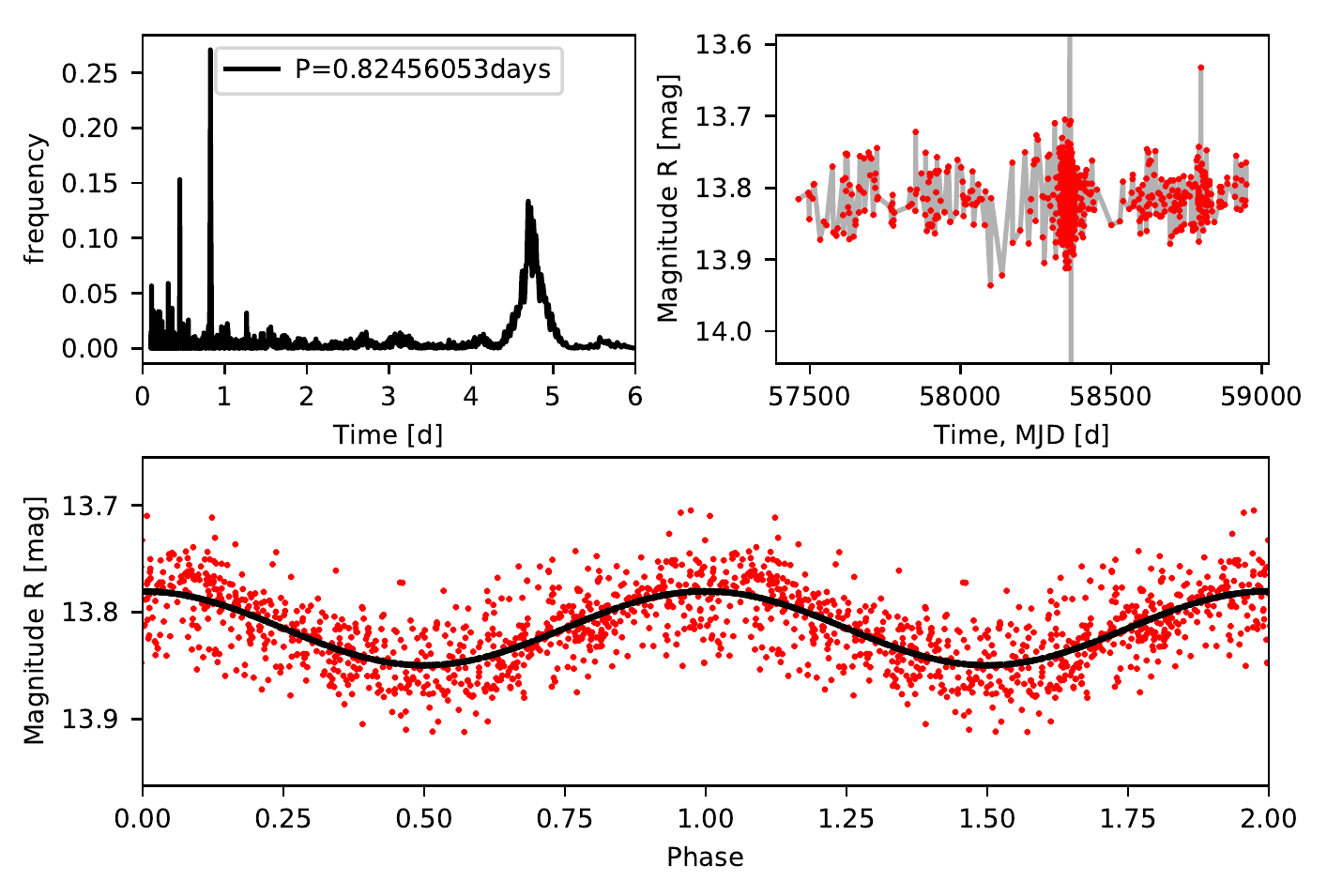} \\
\includegraphics[width=0.99\columnwidth]{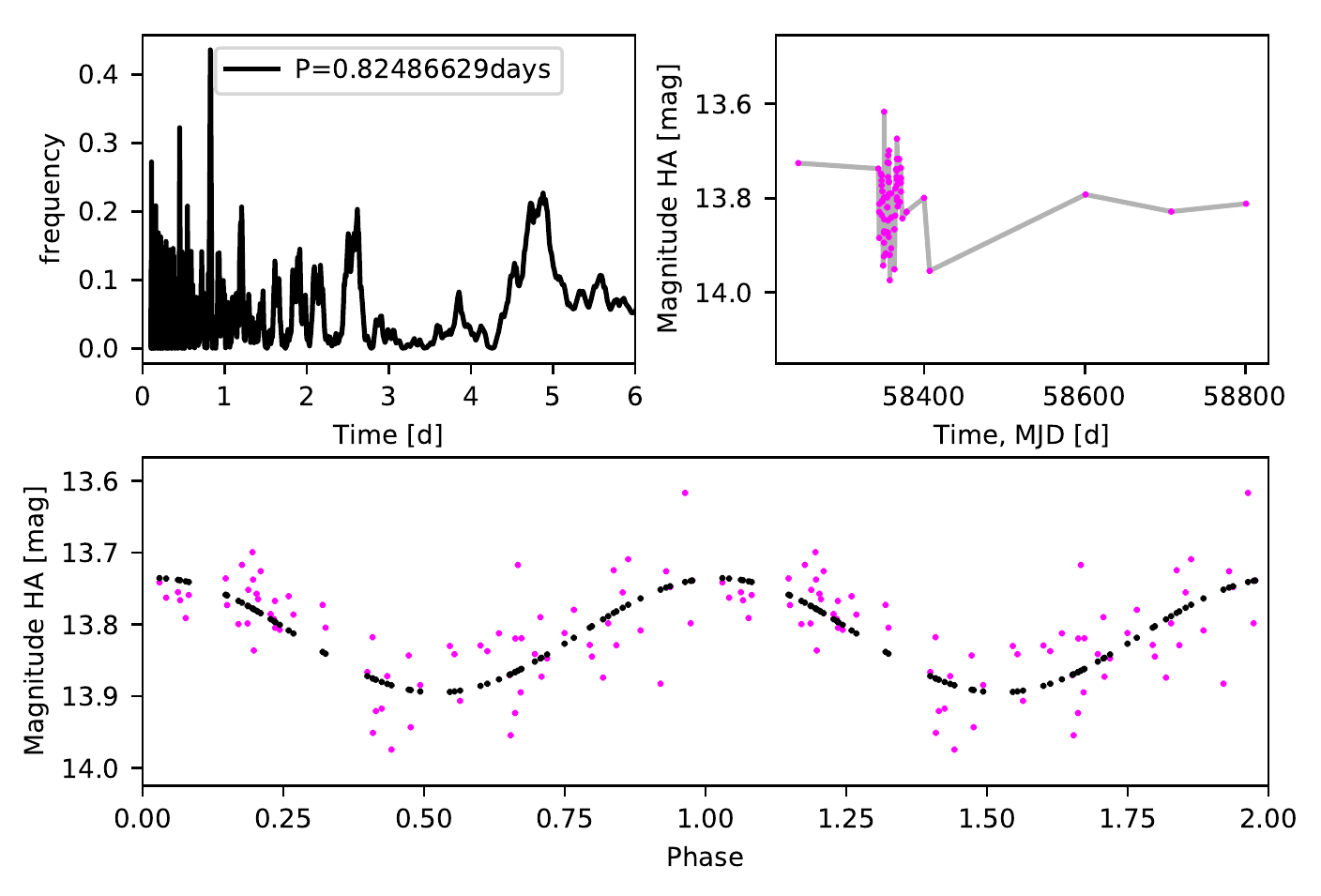} \hfill
\includegraphics[width=0.99\columnwidth]{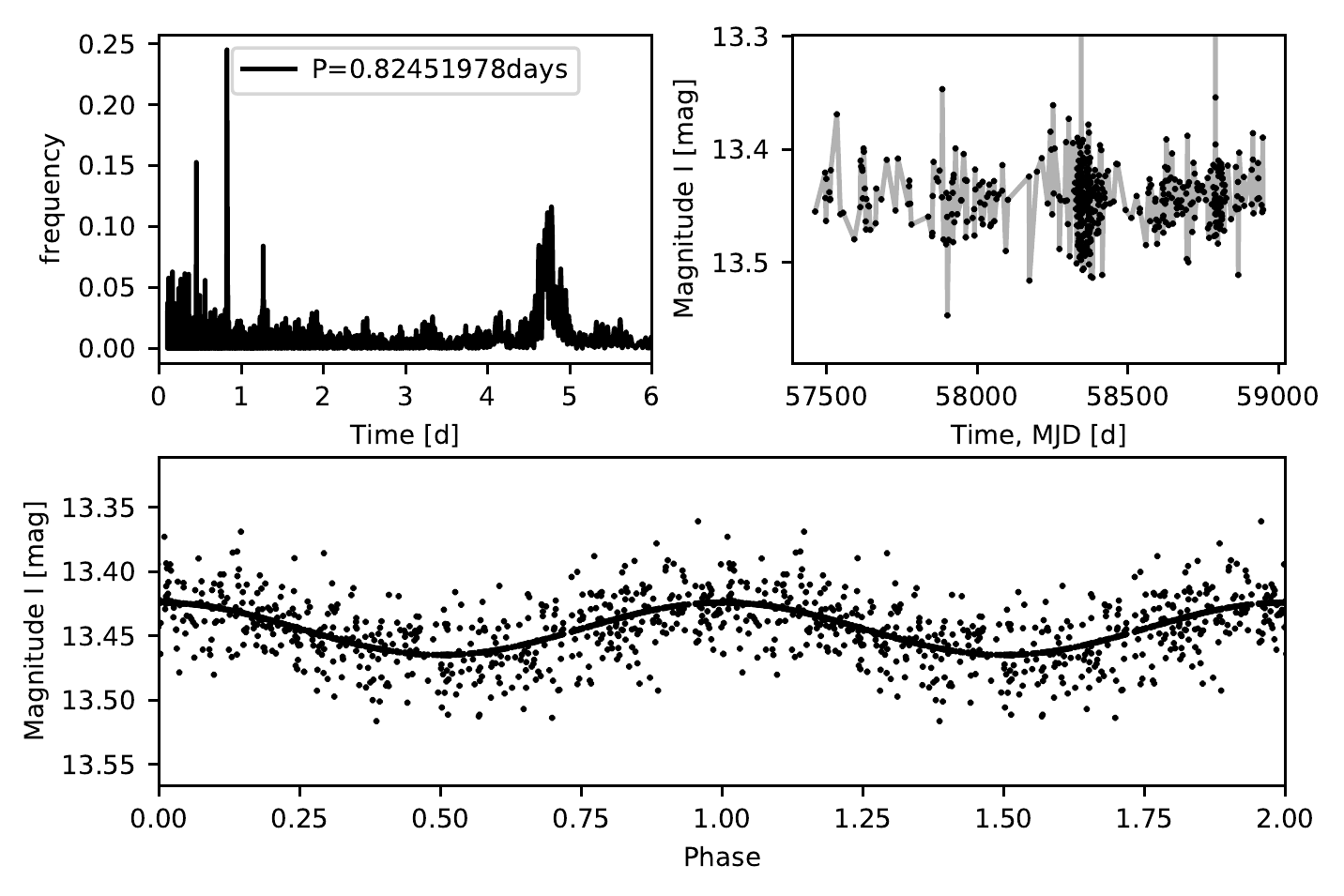} \\
\caption{\label{fig_lombscarcle_all} Lomb-Scargle periodograms, light curves and phased light curves in all filters for the entire data set of \vc.}
\end{figure*}

\bsp	
\label{lastpage}
\end{document}